\documentclass[11pt]{article}

\usepackage{amsmath, amsfonts, amsthm, amssymb, graphicx}
\usepackage{xspace}
\usepackage{epsfig}
\usepackage{psfrag}
\usepackage{hyperref}
\usepackage[dvips]{color}
\numberwithin{equation}{section}

\newcommand{\op}{\ensuremath{\mathcal{O}}\xspace}

\newcommand{\vev}[1]{\ensuremath{\langle #1 \rangle}\xspace}

\def\ie{{\it i.e.\ }}

\let\a=\alpha \let\b=\beta \let\g=\gamma \let\d=\delta \let\e=\epsilon
\let\z=\zeta    \let\k=\kappa
\let\m=\mu \let\n=\nu  \let\r=\rho
\let\s=\sigma \let\t=\tau

    \let\G=\Gamma
\let\del=\partial
\let\cdel=\nabla
\newcommand{\hf}{\frac{1}{2}}
\newcommand{\qt}{\frac{1}{4}}

\def\dalemb#1#2{{\vbox{\hrule height .#2pt
        \hbox{\vrule width.#2pt height#1pt \kern#1pt
                \vrule width.#2pt}
        \hrule height.#2pt}}}

\def\0{{\sst{(0)}}}
\def\1{{\sst{(2)}}}
\def\2{{\sst{(2)}}}
\def\3{{\sst{(3)}}}
\def\4{{\sst{(4)}}}
\def\5{{\sst{(5)}}}
\def\6{{\sst{(6)}}}
\def\7{{\sst{(7)}}}
\def\8{{\sst{(8)}}}

\def\nn{\nonumber}

\newcommand{\be}{\begin{equation}}
\newcommand{\ee}{\end{equation}}
\def\ba{\begin{array}}
\def\ea{\end{array}}

\def\del{\partial}
\def\sst#1{{\scriptscriptstyle #1}}
 
\def\ie{{\it i.e.\ }}

\def\bo1{ \left | B^0 (p^+) \right \rangle}

\def\<{ \langle }
\def\>{ \rangle }

\newcommand{\bea}{\begin{eqnarray}}
\newcommand{\eea}{\end{eqnarray}}

\newcommand{\tr}{{\rm tr} }
\newcommand{\Tr}{{\rm Tr} }

\newcommand{\ams}
{{\it Institute for Theoretical Physics, \\
Valckenierstraat 65, 1018 XE Amsterdam, The Netherlands}\\
{\tt K.Skenderis, M.Taylor, B.C.vanRees@uva.nl}}

\newcommand{\auth}{\large Kostas Skenderis, Marika Taylor and Balt C. van Rees}

\begin{document}

\begin{flushright}
ITF-2009-17
\end{flushright}

\vspace{25pt}

\begin{center}

{\Large \bf  Topologically Massive Gravity \\ and the AdS/CFT Correspondence}

\vspace{20pt}

\auth

\vspace{15pt}


{\ams}

\vspace{20pt}

\begin{abstract}

We set up the AdS/CFT correspondence for topologically massive
gravity (TMG) in three dimensions. The first step in this 
procedure is to determine the appropriate fall off conditions
at infinity. These cannot be fixed a priori as they depend on 
the bulk theory under consideration and
are derived by solving asymptotically the 
non-linear field equations. We discuss in detail the asymptotic
structure of the field equations for TMG, showing that it 
contains leading and subleading logarithms, determine the map between
bulk fields and CFT operators, obtain the appropriate counterterms
needed for holographic renormalization and compute holographically
one- and two-point functions at and away
from  the ``chiral point'' ($\m=1$).  The 2-point functions
at the chiral point are those of a logarithmic CFT (LCFT) with
$c_L=0, c_R=3 l/G_N$ and $b=- 3 l/G_N$, where $b$ is a parameter
characterizing different $c=0$ LCFTs.
The bulk correlators away from the chiral point ($\m \neq 1$) smoothly
limit to the LCFT ones as $ \mu \to 1$. Away from
the chiral point, the CFT contains a state of negative norm
and the expectation value of the energy momentum tensor in that state
is also negative, reflecting a corresponding bulk instability
due to negative energy modes.

\end{abstract}

\end{center}
\newpage
{\small \tableofcontents}
\newpage

\section{Introduction}

Although three-dimensional Einstein gravity is locally trivial, this
is generally no longer the case once higher-derivative terms are added
to the action. The addition of such terms provides the theory with
propagating degrees of freedom, \ie three-dimensional gravitons. The
quantization of such theories therefore appears to give a richer
structure than the Einstein theory, yielding potentially interesting
toy models for higher-dimensional theories of quantum gravity.

Unfortunately, the addition of generic higher-derivative terms to the
Einstein-Hilbert action often gives ghost-like excitations which
render the theory unstable. Recently a renewed interest
has been taken in
the so-called  topologically massive
(cosmological) gravity \cite{Deser:1982vy,Deser:1981wh}, or TMG for short.
This theory consists of
the Einstein-Hilbert action with a negative cosmological
constant plus a gravitational Chern-Simons term
\be
S_{\text{cs}} = \frac{1}{32 \pi G_N \m} \int d^3 x \, \sqrt{-G} \e^{\lambda\m\n} \Big( \G_{\lambda \s}^\r \del_\m \G_{\r \n}^\s + \frac{2}{3}\G_{\lambda \s}^\r \G_{\m \t}^{\s}\G_{\n \r}^\t \Big).
\ee
Although adding a
Chern-Simons term likely leads to instabilities for general values of
the dimensionless parameter $\m$, it was argued in \cite{Li:2008dq}
that the theory
becomes stable and \emph{chiral} when $\m = 1$. At that point, which
we will call the ``chiral point'', all the
left-moving excitations of the theory would become pure gauge and one
would effectively have a right-moving theory.

Other authors however found non-chiral modes at the chiral point,
\cite{Carlip:2008jk,Grumiller:2008qz,Park:2008yy,Grumiller:2008pr,Carlip:2008eq,Carlip:2008qh,Giribet:2008bw,Blagojevic:2008bn}
(see however also \cite{Li:2008yz}). In particular in
\cite{Grumiller:2008qz} a left-moving excitation of the linearized
equations of motion was explicitly written down\footnote{Solutions 
of the non-linear equations of motion exhibiting similar
asymptotic form were presented earlier in 
\cite{AyonBeato:2004fq, AyonBeato:2005qq}.}. From the
transformation properties of the new mode of \cite{Grumiller:2008qz}
under the $(L_0,\bar L_0)$ operators one found a structure typical of
a logarithmic conformal field theory (LCFT) and consequently it was
claimed that the theory with $\m = 1$ was dual to such a theory. Since
LCFTs are not chiral (and not unitary either), this provided a
further argument against the conjecture.

However, near the conformal boundary the new mode does not obey the
same falloff conditions as the other modes. This has led to claims
that one can ignore the new mode by imposing strict `Brown-Henneaux'
\cite{Brown:1986nw}
boundary conditions: the new mode does not satisfy these so it then has to
be discarded and the resulting theory could again be chiral
\cite{Strominger:2008dp}. In \cite{Giribet:2008bw} a non-chiral
mode of the linearized equations of motion, related to that of Grumiller
and Johansson but satisfying the
Brown-Henneaux boundary conditions, was found. However, \cite{Maloney:2009ck}
argued that this mode is not a linearization of a  non-linear
solution. This linearization instability
was further discussed in \cite{Carlip:2009ey}.
On the other hand, in
\cite{Grumiller:2008es,Henneaux:2009pw} it was claimed that the
Brown-Henneaux boundary conditions could be relaxed to incorporate the non-chiral
mode without
destroying the consistency of the theory. At first sight one seems to
be free to choose either set of boundary conditions, supposedly
leading to a different theory for each possibility
\cite{Maloney:2009ck}.

The topologically massive theory admits solutions that are asymptotically AdS so
one can use the AdS/CFT correspondence to analyze the theory. This
is the viewpoint pursued in this paper. One of the
cornerstones of the AdS/CFT correspondence is that the boundary
fields parameterizing
the boundary conditions of the bulk fields are identified with
the sources for the dual operators. It follows that the {\emph{leading}}
boundary behavior must be specified by unconstrained fields, {whereas}
the {\emph{subleading}} radial behavior of the fields is determined
\emph{dynamically} by the equations of motion and should not
be fixed by hand. Putting it differently, the subleading radial
behavior is obtained by finding {\it the most general}
asymptotic solution to the field equations given boundary data.
For theories that admit asymptotically locally AdS solutions
the most general asymptotic solution, which is sometimes called
the ``Fefferman-Graham'' expansion, can always be found by
solving {\it algebraic} equations, see \cite{Skenderis:2002wp}
for a review. We would like to emphasize
that the Fefferman-Graham expansion does {\it not} have a predetermined
form, as is sometimes stated in the literature, but rather the
form of the expansion is dynamically determined.

For theories that admit asymptotically (locally) AdS solutions
finite conserved charges can always be obtained
\cite{Henningson:1998gx,Henningson:1998ey,Balasubramanian:1999re,deHaro:2000xn,
Skenderis:2000in,Papadimitriou:2005ii}
via the formalism of holographic renormalization
\cite{Skenderis:2002wp}. In particular, Ref.
\cite{Papadimitriou:2005ii} provides a first principles proof that
the holographic charges are the correct gravitational conserved charges
for Asymptotically locally AdS spacetimes.
One should contrast the logic here with what is usually done in other papers.
The discussion there starts by selecting fall off conditions for all
fields, for example Brown-Henneaux boundary conditions,
such that interesting known solutions (such as black holes etc.)
are within the allowed class and then it is checked whether these boundary conditions
lead to finite conserved charges. On the other hand, here we start by {\it deriving}
the most general Asymptotically locally AdS boundary conditions.
Finite conserved charges (which satisfy all expected properties)
are guaranteed by the general results of \cite{Papadimitriou:2005ii}.
Note that the finite conserved charges are related to the 1-point function of
the dual energy momentum tensor via the AdS/CFT dictionary. The next
simplest quantities to compute are the 2-point functions of the dual
operators. These are obtained from solutions of the linearized equations
of motion with Dirichlet boundary conditions.

In this paper we develop the AdS/CFT dictionary for topologically massive
gravity. We obtain the most general asymptotic solutions that
are Asymptotically locally AdS and compute the holographic one- and
two-point functions of the theory at and away from the chiral point.
One new feature in this case is that the field equations are third order 
in derivatives. Ordinarily higher derivative terms are treated as
perturbative corrections to two derivative actions and as such 
they do not change the usual AdS/CFT set-up. In the case of TMG, however,
we need to treat the Einstein and Chern-Simons terms on equal footing.
The fact that the field equation is third order implies that 
there is an additional piece of boundary data to be specified. This means that 
we can fix both a boundary metric (or more precisely, a conformal class)
and (part of) the extrinsic curvature. The boundary metric acts as 
a source for the boundary stress energy tensor, while the field 
parametrizing the boundary condition for the extrinsic curvature 
is a source for a new operator. It turns out that this operator 
is irrelevant when $\m >1$ and it becomes the logarithmic partner of 
the stress energy tensor as $\mu \to 1$.

The asymptotic expansion at $\m=1$ contains the subleading log
piece found earlier in  \cite{Grumiller:2008qz}. The coefficient of this
term corresponds to the 1-point function of 
the logarithmic partner of the energy momentum tensor. As 
this operator is obtained as a limit of an irrelevant
operator, its source (as usual) should be treated
perturbatively.  This source, which is the above mentioned boundary 
condition for the extrinsic curvature, appears as
the coefficient of a {\it leading order log} term in the
solution to the linearized equations of motion (not to be confused
with the subleading log of \cite{Grumiller:2008qz} which relates
to the 1-point function of this operator). The results for the two-point 
functions at $\m=1$ completely
agree with LCFT expectations and the results away from $\m=1$ smoothly
limit to the $\m=1$ results. Bulk instabilities when $\m \neq 1$
due to negative energy modes
also neatly map to properties of the boundary theory, namely negative norm
states and correspondingly
negativity of the expectation value of the energy momentum tensor
in these states.

The remainder of the paper is structured as follows. After
discussing some conventions and giving the equations of motion, we review in
section \ref{sec:aladsspacetimes} the standard AdS/CFT dictionary,
in particular the definition of Asymptotically locally AdS spacetimes, and
point out several subtleties which will be crucial in its application
to TMG. In section \ref{sec:asymptoticanalysis} we analyze the
asymptotic structure of the bulk solutions for $\m = 1$. We compute
the on-shell action, discuss its divergences and the holographic
renormalization which enables us to concretely formulate the
holographic dictionary. The holographic one point functions satisfy anomalous Ward identities
whose interpretation is discussed in section \ref{sec:anomaly}.
Section \ref{sec:linearizedanalysis} concerns
linearized analysis which is used to compute holographically one- and
two-point functions for $\m =1$. We then repeat this analysis for
general $\mu$ in section \ref{sec:linearizedgeneralmu}. We end with a
short summary and an outlook. Various appendices contain
computational details as well as a discussion of some relevant aspects of
logarithmic CFTs.

\section{Setup and equations of motion}
The bulk part of the action has the form:
\be
\label{eq:bulkaction}
\begin{split}
S&= \frac{1}{16 \pi G_N} \int d^3 x \, \sqrt{-G}(R - 2 \Lambda) \\ &\qquad + \frac{1}{32 \pi G_N \m} \int d^3 x \, \sqrt{-G} \e^{\lambda\m\n} \Big( \G_{\lambda \s}^\r \del_\m \G_{\r \n}^\s + \frac{2}{3}\G_{\lambda \s}^\r \G_{\m \t}^{\s}\G_{\n \r}^\t \Big),
\end{split}
\ee
where we use the covariant $\e$-symbol such that $\sqrt{-G}\epsilon^{012} = 1$ with $x^2$ the radial direction denoted $\rho$ below. We set $\Lambda = -1$ below. We use the following conventions for the curvatures:
\be
R_{\m \n \r}^{\phantom{\m \n \r}\s} = \del_\n \G_{\m \r}^\s + \G_{\m \r}^{\lambda}\Gamma_{\n\lambda}^\s - (\m \leftrightarrow \n), \qquad \qquad R_{\m \r} = R_{\m \s\r}^{\phantom{\m \s \r}\s}.
\ee
All Greek indices run over three dimensions, all Latin indices over two dimensions. In three dimensions the Weyl tensor vanishes identically, which means that:
\be
\label{eq:riemann3d}
R_{\m\n\r\s} = G_{\m\r}R_{\s\n} - G_{\n \r} R_{\s \m} - \hf R G_{\m\r}G_{\s\n} - (\r \leftrightarrow \s ).
\ee
The equation of motion derived from \eqref{eq:bulkaction} becomes:
\be
\label{eq:eom}
R_{\m \n} - \hf G_{\m \n}R - G_{\m \n} + \frac{1}{\m}C_{\m \n} = 0,
\ee
with $C_{\m \n}$ the Cotton tensor:
\be \label{cotton}
C_{\m \n} = \epsilon_\m^{\phantom{\m}\a\b}\cdel_\a(R_{\b \n} - \qt R G_{\b \n}).
\ee
Using \eqref{eq:riemann3d} we find that the Bianchi identity becomes:
\be
C_{\m \n} - C_{\n \m}= 0\,.
\ee
The last term in the r.h.s. of (\ref{cotton}) is totally antisymmetric in $\m$ and $\n$ and therefore merely subtracts the antisymmetric piece from the
first term  in the r.h.s. of (\ref{cotton}). We alternatively have:
\be
C_{\m \n} = \frac{1}{2}\Big(\epsilon_\m^{\phantom{\m}\r\s}\cdel_\r R_{\s \n} + \epsilon_\n^{\phantom{\n}\r\s}\cdel_\r R_{\s \m}\Big).
\ee
It is not hard to verify that
\be
C_\m^\m = 0, \qquad \qquad \cdel_\m C^{\m \n} = 0\,.
\ee
Taking the trace of \eqref{eq:eom} we therefore find that:
\be \label{constR}
R = -6,
\ee
independent of $\mu$. Substituting this back, we find:
\be
\label{eq:simpleeom}
R_{\m \n} + 2 G_{\m \n} + \frac{1}{\m}\epsilon_\m^{\phantom{\m}\r\s}\cdel_\r R_{\s \n} = 0,
\ee
from which we also obtain that any solution to the Einstein equations has $R_{\m \n} = -2 G_{\m \n}$ and is a solution to these equations as well.

\section{Asymptotically AdS spacetimes and holography}
\label{sec:aladsspacetimes}

In this section we will explain what
Asymptotically (locally) AdS, or A(l)AdS spacetimes
are and their role in the AdS/CFT correspondence.
Reviews of the mathematical aspects discussed here
can be found in \cite{Graham:1999jg,Anderson:2004yi}.
After introductory comments that are
generally applicable,
we highlight two aspects of the framework that
will be important for its application to TMG, namely irrelevant
deformations and higher-derivative terms.

\subsection{Conformally compact manifolds}
First of all, we define a $D$-dimensional \emph{conformally compact}
manifold-with-metric $(M,G)$ as follows. Let $M$ be the interior of a
manifold $\bar M$ with boundary $\del M$.\footnote{For the purpose of this
introduction we take the manifold to be Euclidean (so in particular
$\del M$ does not contain initial and final
hypersurfaces). The Lorentzian case can be dealt with using the formalism of
\cite{us,Skenderis:2008dg}.} Suppose there exists a smooth, non-negative
\emph{defining function} $z$ on $\bar M$ such that $z(\del M) = 0$, $d
z (\del M) \neq 0$ and the metric
\be
\tilde G = z^2 G
\ee
extends smoothly to a {\it non-degenerate} metric on $\bar M$. We then say
that $(M,G)$ is conformally compact and the choice of a defining
function determines a \emph{conformal compactification} of $(M,G)$.

The metric $\tilde G$ induces a regular metric $g_{(0)}$ on $\del
M$. This metric depends on the defining function, as picking a
different defining function Weyl rescales $g_{(0)}$. It follows that
the pair $(M,G)$ determine a conformal structure (denoted $[g_{(0)}]$)
at $\del M$. We call $(\partial M,[g_{(0)}])$ the \emph{conformal
  infinity} or \emph{conformal boundary} of $(M,G)$.
This construction is same as the Penrose method of compactifying
spacetime by introducting conformal infinity.

If we compute the Riemann tensor of $G$, we find that near $\del M$ it has the form:
\be
R_{\m \n \r\s} = - \tilde G^{\k \lambda} \cdel_\k z \cdel_\lambda z (G_{\m \r}G_{\n \s} - G_{\n \r}G_{\m \s}) + O(z^{-3}).
\ee
Notice that the leading term is order $z^{-4}$ as $G$ is order $z^{-2}$. Taking its trace we obtain that:
\be
R = - D (D-1) \tilde G^{\k \lambda} \cdel_\k z \cdel_\lambda z + O(z).
\ee
We see that for a spacetime with constant negative curvature,
\be
R = - D(D-1),
\ee
and thus we find to leading order:
\be
\tilde G^{\k \lambda} \cdel_\k z \cdel_\lambda z = 1.
\ee
The Riemann curvature of such a metric thus approaches that of AdS
space with cosmological constant $\Lambda = - (D-1)(D-2)/2$, for which
$R_{\m \n \r\s} = - D(D-1) (G_{\m \r}G_{\n \s} - G_{\n \r}G_{\m \s})$
holds exactly. A conformally compact manifold whose metric also
satisfies $R = -D(D-1)$ is therefore also called an
\emph{Asymptotically locally AdS} manifold. Notice that we added the
word `local' because we have not put any requirements on
global issues like the
topology of $\del M$, which may very well be different from the sphere
at conformal infinity of (Euclidean) AdS.

\subsection{Fefferman-Graham metric}
\label{sec:fgmetric}
A main result of Fefferman and Graham \cite{FeffermanGraham} is that
in a finite neighborhood of $\partial M$,
the metric of an AlAdS spacetime can always be cast in the form:
\be
\label{eq:fgmetricz}
ds^2 = z^{-2}(dz^2 + g_{ij}dx^i dx^j),
\ee
where the conformal boundary is at $z=0$ and the metric $g$ is a
regular metric at $\del M$, which we can write as:
\be
\label{eq:g0}
g_{ij}(x^k,z) = g_{(0)}(x^k) + \ldots,
\ee
where the dots represent terms that vanish as $z \to 0$.
The coordinates in (\ref{eq:fgmetricz}) are Gaussian normal coordinates
centered at $\partial M$.

The specific form of the subleading terms, including the radial power
where the first subleading terms appears, depends on the bulk theory
under question and is not fixed a priori. For example, for Einstein
gravity in $(d+1)$ dimensions the expansion reads
\be
\label{eq:fgexpansiong}
g_{ij} = g_{(0)ij} + z^2 g_{(2)ij} + \cdots + z^d (g_{(d)ij}
+ h_{(d)ij} \log(z))
+ \cdots
\ee
The fact that the subleading term starts at order $z^2$ is
specific to pure Einstein gravity. For example, $3d$ Einstein gravity
coupled to matter can have the first subleading term appearing
at order $z$, see \cite{Berg:2001ty} for an example.
The logarithmic term $h_{(d)}$ appears
in Einstein gravity when $d$ is an even integer greater than 2.
This coefficient is given by
the metric variation of the conformal anomaly \cite{deHaro:2000xn}.
This fact immediately explains why there is no such coefficient
in Einstein gravity when $d=2$: in this case the conformal anomaly is
given by a topological invariant and therefore its variation w.r.t.
the metric vanishes. As soon as the bulk action contains additional fields
the expansion will be modified accordingly
\cite{deHaro:2000xn,Bianchi:2001de,Berg:2001ty,Bianchi:2001kw}.
For example, the asymptotic solution
for three dimensional Einstein gravity coupled to a free massless
scalar field is of the form (\ref{eq:fgexpansiong}) with a non-zero
$h_{(2)}$ coefficient, see equation (5.25) of \cite{deHaro:2000xn}\footnote{
Ref. \cite{Kanitscheider:2006zf}, appendix E, contains an example of
3d gravity coupled to scalars
with $\log^2$ terms in the asymptotic expansion.}.
Note that the log term found in \cite{Grumiller:2008qz}
is precisely of this form. From this perspective the appearance
of such a term in the asymptotic expansion of TMG is certainly not
surprising.

What is universal in this discussion is the structure of these expansions.
The subleading
coefficients are determined locally in terms of $g_{(0)}$ by solving
asymptotically the field equations. This procedure leads to algebraic
equations that can be readily solved.
On the other hand, $g_{(d)}$ is not locally determined by $g_{(0)}$
but rather by global constraints like regularity of the bulk metric in
the interior of $M$. This term is related to the 1-point function of
$T_{ij}$.

To repeat, according to the
standard AdS/CFT dictionary the allowed subleading terms in expansions
like \eqref{eq:fgexpansiong} (and \eqref{eq:expansionphi} below) are
determined by the equations of motion rather than fixed by hand. As
long as $g$ is regular for $z = 0$ and therefore of the form
\eqref{eq:g0}, the aforementioned AlAdS properties of $(M,G)$ are
unchanged. In the context of TMG this in particular implies 
that we allow the logarithmic mode found in \cite{Grumiller:2008qz}.

\subsection{Boundary conditions and dual sources}

According to the AdS/CFT dictionary \cite{Gubser:1998bc,Witten:1998qj}, the coefficients of the leading terms in the radial expansion of the metric and the various matter fields are sources for corresponding gauge-invariant operators in the CFT. For example, $g_{(0)}$ specifies a boundary metric which becomes the source for the energy-momentum tensor of the boundary theory. Similarly, a bulk scalar field $\Phi$ of mass $m$ has the allowed asymptotic behavior:
\be
\label{eq:expansionphi}
\Phi = \phi_{(0)} z^{d - \Delta} + \ldots + \phi_{(2\Delta -d)}z^{\Delta} + \ldots
\ee
with $m^2 = \Delta (\Delta -d)$. We then interpret the leading term $\phi_{(0)}$ as the source for a scalar operator $\op$ of scaling dimension $\Delta$ dual to $\Phi$.

In field theory, one computes the partition function as a functional
of sources and the same story applies in AdS/CFT. The sources like
$\phi_{(0)}$ and $g_{(0)}$ determine the asymptotic (Dirichlet) values
of a bulk solution to the equation of motion. The aim is now to find
this bulk solution and subsequently compute its on-shell action. Since
the solution of the equations of motion is a function of $\phi_{(0)}$
and $g_{(0)}$, so is the corresponding on-shell action.
However,
the naive action is always infinite (for example, the Einstein-Hilbert
term is proportional to the volume of spacetime which always diverges
for an AlAdS spacetime). We therefore need to regularize and then
renormalize the computation of the on-shell action. This
\emph{holographic} renormalization of the on-shell action depends
crucially on the asymptotic properties of the metric (which in
our case is AlAdS) and this is the place where the above framework
finds a practical application.

Holographic renormalization is implemented as follows, see
\cite{Skenderis:2002wp} for a more complete discussion. One first puts
the boundary of the spacetime at finite $z_0$ rather than at $z = 0$
and then evaluates the on-shell action for this regulated
solution. One finds divergences as $z_0 \to 0$ which can however be
cancelled by adding local counterterms to the action. To maintain
covariance, these counterterms should be functionals of the induced
metric and other fields on the slice given by $z = z_0$. Adding then
the counterterms to the on-shell action, one finds that the total
action is finite as $z_0 \to 0$.

Once the on-shell action is renormalized and finite, one can compute
one-point functions in the presence of sources by functionally
differentiating the renormalized on-shell action with respect to the
sources like $g_{(0)}$ and $\phi_{(0)}$. These one-point functions involve the nonlocally
determined pieces called $g_{(d)ij}$ and $\phi_{(2\Delta - d)}$
and in general contain also local terms, some of which are
related to anomalies and others that are scheme dependent. One can obtain higher-point
functions by taking further derivatives of the one-point
functions and the  local terms lead to contact terms in $n$-point functions.

Notice that the counterterms are also necessary for the appropriate
variational principle to hold: for AlAdS spacetimes
one fixes $g_{(0)}$ (or rather its conformal class) instead of the
induced boundary metric $g/z_0^2$ which would diverge as $z_0 \to
0$. This is discussed in detail in \cite{Papadimitriou:2005ii}.

\subsection{Sources for irrelevant operators}
\label{sec:aladsholography}

The fact that an asymptotically AdS  metric
becomes that of AdS near conformal infinity is dual to the
statement that the boundary theory becomes conformal at high
energies. Asymptotically AdS metrics describe
relevant deformations of the CFT and/or vevs in the boundary theory.

On the other hand, one may also attempt to switch on sources for
irrelevant operators. Such deformations are for example necessary to
compute correlation functions of irrelevant operators, as these are
obtained by functionally differentiating the on-shell action with
respect to these sources. Switching on these sources spoils the
conformal UV behavior of the field theory. Correspondingly, the bulk
solutions will no longer be AlAdS and the usual AdS/CFT dictionary
would break down. In particular, the usual counterterms no longer
suffice to make the on-shell action finite, completely analogous to
the nonrenormalizability of the field theory with such sources.

A consistent perturbative approach may however be set up by treating
the sources for irrelevant operators as infinitesimal \cite{deHaro:2000xn}.
In the bulk,
this means that one starts from an AlAdS solution and computes the
bulk solution and the on-shell action to any given order $n$ in the
sources. This approximation allows for the computation of $n$-point
functions of the irrelevant operator in any given state dual to the
background AlAdS solution. We will see a concrete example worked out
below.

\subsection{Higher-derivative terms}
\label{sec:higherderivative}
Higher-derivative terms in the bulk action are usually treated perturbatively and in that case do not directly lead to a change in the setup described above. However, for TMG we cannot afford to treat these terms as perturbations as we want to study the complete theory around $\m = 1$. The solution to the bulk equations of motion is then generally no longer fixed by the specification of Dirichlet data alone and some extra boundary data is needed; for example the $z$-derivatives of the metric $g_{ij}$ at the boundary. Correspondingly, the on-shell action depends on these boundary data as well. We shall see below that this is precisely what happens for TMG.

Extending the usual AdS/CFT logic, we interpret the new boundary data as a new source for another operator in the field theory. Functionally differentiating the on-shell action with respect to this new boundary data then yields correlation functions of this new operator. To make contact with earlier results, notice that for TMG this operator creates the massive graviton states in the bulk and for $\m = 1$ it creates the logarithmic solution found in \cite{Grumiller:2008qz}. One may say that these spaces have only a single operator insertion in the infinite past.

It turns out that this new operator is irrelevant for $\m > 1$, as for $\m \geq 1$ we find that switching on the corresponding source spoils the AlAdS properties of the spacetime. Following the discussion of the previous subsection, we therefore will have to treat the source as infinitesimal and approach the problem perturbatively to a given order in the source. This is precisely what we will do in section \ref{sec:linearizedactions}.

\section{Asymptotic analysis for $\m =1$}
\label{sec:asymptoticanalysis}

In this section we return to TMG and carry out an asymptotic analysis
of the equations of motion \eqref{eq:eom} in the Fefferman-Graham
coordinate system. Note that because of (\ref{constR}) all conformally
compact solutions of this theory are asymptotically locally AdS.
However, not all solution of TMG are conformally compact.
For example, the `warped' solutions of \cite{Anninos:2008fx}
have a degenerate boundary metric, as is demonstrated in appendix
\ref{app:warped}, and thus they are not conformally compact.
In this section we restrict to the AlAdS case.
We compute the on-shell action, discuss the
variational principle in detail and demonstrate how one
holographically computes one-point functions in the CFT. As indicated
in the previous section, we will find irrelevant operators and
therefore the complete holographic renormalization of the on-shell
action has to be done perturbatively. This is postponed until the next
section, where we will renormalize the action to second order in the
perturbations.

Although this and the next section focus on the case $\m = 1$, $\m$ is sometimes reinstated for later 
convenience.

\subsection{Fefferman-Graham equations of motion}
\label{sec:fgeom}
Following the discussion in section \ref{sec:fgmetric}, we take the metric to be of the form:
\be
\label{eq:fgmetric}
ds^2 = \frac{d\r^2}{4\r^2} + \frac{1}{\r} g_{ij}(x,\r)dx^i dx^j
\ee
where we defined $\r = z^2$. As should be clear from the previous
section, this form of the metric is not an ansatz but it is
a direct consequence of the AlAdS property of the spacetime.
In other words, the metric of any  AlAdS spacetime can be brought to
this form near the conformal boundary.
In this coordinate system the equations
of motion \eqref{eq:eom} take the following form. For the component
equations we find:

\be
\label{eq:eomfgform}
\begin{split}
&- \hf \tr(g^{-1}g'') + \qt \tr(g^{-1}g' g^{-1}g') + \frac{1}{4\m}\e^{ij}\Big( \cdel_i \cdel^k g'_{kj} + 2 \r(g'' g^{-1}g')_{ji} \Big) = 0,\\
& \Big(\frac{1}{2}\tr(g^{-1}g'') - \qt [\tr(g^{-1}g')]^2 \Big)g_{ij}  - g''_{ij} + \hf g'_{ij} \tr(g^{-1}g')  \\&\qquad  + \frac{1}{\m}\e_i^{\phantom{i} k}\Big\{ \qt \cdel_k \cdel^m g'_{mj} + \qt \cdel_j \cdel^m g'_{mk} - \hf \cdel_k \cdel_j[\tr(g^{-1}g')] + 2 \r g'''_{jk} + \\ &\qquad  g''_{kj}[3 -\frac{3}{2} \r \tr(g^{-1}g')] + g'_{kj}\Big(- \frac{3}{2}\tr(g^{-1}g') + \frac{3}{4} \r [\tr(g^{-1}g')]^2 \\&\qquad - \frac{7}{2} \r \tr(g^{-1}g'') + \frac{7}{4}\r \tr(g^{-1}g'g^{-1}g')\Big)
\Big\} + i \leftrightarrow j = 0,\\
& (g^{kj} - \m \e^{kj}) \cdel_k g'_{ij} - \cdel_i \Big( \tr(g^{-1}g') + \hf \r \tr(g^{-1}g'g^{-1}g') - \r [\tr(g^{-1}g')]^2\Big) \\
& \qquad + 2 \r \cdel^n \Big(g''_{in} - \tr(g^{-1}g') g'_{in}\Big) + \r (g^{-1}g')^k_i \cdel^l g'_{kl} = 0\,,
\end{split}
\ee
whereas the trace equation $R= -6$ becomes:
\be
\label{eq:R}
- 4 \r \tr(g^{-1}g'') + 3 \r \tr(g^{-1}g' g^{-1}g') - \r [\tr(g^{-1}g')]^2 + R(g) + 2 \tr(g^{-1}g') = 0.
\ee
A prime denotes a derivative with respect to $\r$. The derivation of these equations is given in appendix \ref{sec:appeom}.

\subsection{Asymptotic solution}
\label{sec:asymptoticsoln}
Rather than the usual asymptotic behavior $\lim_{\r \to 0} g_{ij}(\r,x^k) = g_{(0)ij}(x^k)$, the equations of motion for $\mu = 1$ also allow
leading log asymptotics for $g_{ij}$. We therefore substitute the expansion
\be
\label{eq:fgexpansion}
g_{ij} = b_{(0)ij} \log(\r) + g_{(0)ij} + b_{(2)ij} \r \log(\r) + g_{(2)ij} + \ldots
\ee
into the equations of motion. The subleading logarithmic term
$b_{(2)ij}$ in this expansion is the mode considered in
\cite{Grumiller:2008qz}. The leading logarithmic term $b_{(0)ij}$, on
the other hand, changes the asymptotic structure of the spacetime and
it is no longer AlAdS. Following the discussion in section
\ref{sec:aladsholography}, we will treat $b_{(0)ij}$ to be
infinitesimal and work perturbatively in $b_{(0)ij}$. As we will be
interested in two-point functions around a background with $b_{(0)ij}
= 0$, it suffices to retain only terms linear in $b_{(0)ij}$ in the
equations that follow.

Under these conditions we find:
\be
\begin{split}
g'_{ij} &= \frac{b_{(0)ij}}{\r} + b_{(2)ij} \log(\r) + b_{(2)ij} +  g_{(2)ij} + \ldots,\\
g''_{ij} &= -\frac{b_{(0)ij}}{\r^2} + \frac{b_{(2)ij}}{\r} + \ldots, \\
g'''_{ij} &= \frac{2 b_{(0)ij}}{\r^3} - \frac{b_{(2)ij}}{\r^2} + \ldots, \\
g^{ij} &= g_{(0)}^{ij} - b_{(0)}^{ij}\log(\r) - b^{ij}_{(2)}\r \log (\r) - \r g_{(2)}^{ij} + \op(b_{(0)}) + \ldots,
\end{split}
\ee
where in the last line indices are raised with $g_{(0)}$ and the $\op(b_{(0)})$ terms are of the form $b^i_{(2)k} b^{kj}_{(0)} \r \log^2(\r) + g_{(2)k}^i b^{kj}_{(0)} \r \log(\r)$, but will never be needed in what follows.

Substituting this expansion in the equations of motion \eqref{eq:eomfgform} and \eqref{eq:R}, we find the following. To leading order we find both from the $(\r\r)$ equation as well as from the $R$ equation that:
\be
\tr(b_{(0)}) = 0.
\ee
Notice that traces are now implicitly taken using $g_{(0)}$, that is $\tr(b_{(0)}) \equiv g_{(0)}^{ij}b_{(0)ij}$. Also, in this subsection the $\e$-symbol and covariant derivatives are defined using $g_{(0)}$. From the $(ij)$ equation we find that:
\be
P_i^k b_{(0)kj} = 0,
\ee
where we define the projection operators:
\be
P_i^k \equiv \hf(\delta_i^k + \e_i^{\phantom{i}k}), \qquad \qquad \bar P_i^k \equiv \hf(\delta_i^k - \e_i^{\phantom{i}k}),
\ee
and we obtain no new constraint from the $(\r i)$ equation at leading order.

At subleading order we encounter various log terms. From the $R$ equation we find at order $\log^2(\r)$ that
\be
\label{eq:trhb}
\tr(b_{(2)} g_{(0)}^{-1} b_{(0)}) = 0
\ee
and at order $\log(\r)$ we then find:
\be
\label{eq:trh}
- 2 \tr(b_{(0)} g_{(0)}^{-1} g_{(2)}) + 2 \tr(b_{(2)}) + \tilde R[b_{(0)}] = 0,
\ee
with $\tilde R[b_{(0)}]$ the linearized curvature:
\be
R[g] = R[g_{(0)}] + \log(\r) \tilde R[b_{(0)}]  + \ldots,
\ee
which can be more explicitly written as:
\be
\tilde R[b_{(0)}] = \cdel^i \cdel^j b_{(0)ij},
\ee
where we used the properties of $b_{(0)ij}$ found at leading order. At subleading order in the $(\r\r)$ equation we again obtain \eqref{eq:trhb} and \eqref{eq:trh}. At order one in the $R$ equation we obtain:
\be
\label{eq:trg2}
- 2 \tr(b_{(2)}) + 2 \tr(g_{(2)}) + R[g_{(0)}] = 0.
\ee
For the $(ij)$ equation the subleading terms at order $\log(\r)/\r$ give
\be
(b_{(0)} g_{(0)}^{-1}b_{(2)})_{ij} + (b_{(2)} g_{(0)}^{-1} b_{(0)})_{ij} = 0,
\ee
and at order $1/\r$ we obtain:
\be
\label{eq:constrainth}
\bar P_i^k b_{(2)kj} = \hf(b_{(2)ij} - \e_i^{\phantom{i}k} b_{(2)kj})  = \op(b_{(0)ij}),
\ee
where the right-hand side is an expression linear in $b_{(0)ij}$ that we will not need below.

For the $(\r i)$ equation, we find at subleading order that:
\be
\bar P_i^k \big( \cdel^j g_{(2)jk} + \hf \cdel_k R[g_{(0)}]\big) = \cdel^l b_{(2)li} + \op(b_{(0)}).
\ee
We may apply \eqref{eq:constrainth} to rewrite schematically $b_{(2)ij} \to P_i^k b_{(2)kj} + \op(b_{(0)})$. Since $P_i^k$ and $\bar P_i^k$ are projection operators onto orthogonal subspaces we can split this equation into:
\be
\label{eq:consg2}
\bar P_i^k \big( \cdel^j g_{(2)jk} + \hf \cdel_k R[g_{(0)}]\big) = \op(b_{(0)}), \qquad \qquad \cdel^l b_{(2)li} = \op(b_{(0)}).
\ee
If $b_{(0)ij} = 0$ then the first of these equations agrees with \cite{Solodukhin:2005ah}.

\subsection{On-shell action}
In this section we will write the on-shell action in Fefferman-Graham coordinates and analyze the divergences obtained by substituting the expansion \eqref{eq:fgexpansion}.

We begin by computing the on-shell value of the Chern-Simons part of the action,
\be
\begin{split}
I_{\text{cs}} &= \frac{1}{32 \pi G_N \m} \int d^3 x \, \sqrt{-G}\e^{\lambda\m\n}\Big( \G_{\lambda \s}^\r \del_\m \G_{\r \n}^\s + \frac{2}{3}\G_{\lambda \s}^\r \G_{\m \t}^{\s}\G_{\n \r}^\t\Big),
\end{split}
\ee
in Fefferman-Graham coordinates. Observing that the $\e$-symbol
implies that only one of the indices $\lambda$, $\m$ or $\n$ can be the
radial direction, we can directly write out the various terms. Using
then \eqref{eq:connections} and \eqref{eq:riemanns} from appendix
\ref{sec:appeom} we find that many terms cancel due to the
antisymmetry of $\e^{ij}$ and we are left with:
\be
\label{eq:onshellcs}
\frac{1}{32 \pi G_N \m}  \int d^3 x \sqrt{-g}\e^{ij} \Big(2 \r (g'g^{-1}g'')_{ij} - \G^{a}_{ib} \partial_\r \G_{aj}^b \Big),
\ee
where the connection coefficients and $\e$ tensor are now those
associated with $g_{ij}$. Substituting \eqref{eq:fgexpansion}, it
is not hard to verify that this action is finite for $\r_0 \to 0$ if
$b_{(0)ij} = 0$, but there are log divergences if $b_{(0)ij}$ is
nonzero.

For the Einstein-Hilbert action, the variational principle can be made
well-defined for Dirichlet boundary conditions at a \emph{finite}
radial distance
by the addition of the Gibbons-Hawking term. In our
conventions, this means that the Einstein part of the action is given
by:
\be
\label{eq:einsteinaction}
I_{\text{gr}} = \frac{1}{16\pi G_N}\int d^3 x \sqrt{-G}(R - 2 \Lambda) + \frac{1}{8 \pi G_N} \int d^2 x \sqrt{-\gamma}K \, ,
\ee
where $\gamma_{ij} = g_{ij}/\r$ is the induced metric on the cutoff surface $\r = \r_0$, which is kept fixed in the variational problem. Furthermore, $K$ is the trace of the extrinsic curvature of this surface, which is defined using the outward
pointing unit normal $n_\m dx^\m = - d\r/(2\r)$.

This variational problem becomes ill-posed as
$\r_0 \to 0$, since the induced metric $\gamma$ diverges in this limit.
What one should instead keep fixed is the conformal class of $\gamma$
(or $g_{(0)}$ after taking into account the issues related to the
conformal anomaly) \cite{Papadimitriou:2005ii}. This requires
introducing additional boundary terms. These boundary terms not
only make the variational problem well-posed but also make the
on-shell action finite as $\r_0 \to 0$. In
particular, for the pure Einstein theory the counterterm action is
\be
\label{eq:ctaction}
I_\text{ct} = \frac{1}{8\pi G_N} \int d^2 x \sqrt{-\gamma} \Big(- 1  + \qt R [\gamma] \log (\r_0) \Big).
\ee
Substituting the Fefferman-Graham form of the metric we find:
\be
\begin{split}
I_{\text{gr}} &= - \frac{1}{16\pi G_N} \int d^3 x \frac{2}{\rho^2}\sqrt{-g} + \frac{1}{16 \pi G_N}\int d^2 x \frac{1}{\r}\sqrt{-g} ( 4 - 2 \r \tr(g^{-1}g')),\\
I_{\text{ct}} &= \frac{1}{8\pi G_N}  \int d^2 x \sqrt{-g} \Big(- \frac{1}{\r_0}
  + \qt R [g] \log (\r_0) \Big).
\end{split}
\ee
We may now substitute the radial expansion \eqref{eq:fgexpansion} for $g_{ij}$ and find the same behavior as for the Chern-Simons part: the action $I_{\text{gr}} + I_{\text{ct}}$ is finite when $b_{(0)ij} = 0$ but diverges otherwise.

We now define the following combined action:
\be
\label{eq:itot}
I_{\text{c}} = I_{\text{gr}} + I_{\text{cs}} + I_{\text{ct}},
\ee
which we emphasize is finite only as long as $b_{(0)ij}$ vanishes and needs to be supplemented with additional boundary counterterms otherwise. As we explained in section \ref{sec:aladsspacetimes}, this will be done perturbatively up to the required order in $b_{(0)ij}$. We will do an explicit analysis to second order in section \ref{sec:linearizedanalysis}, but first we discuss the variational principle and the computation of the one-point functions in general terms.


\subsubsection{Variational principle}
In this subsection we compute the variation of the combined action $I_{\text{c}}$ defined in \eqref{eq:itot}, which will be needed below in the
holographic computation of boundary correlation functions.

First of all, the variation of the Einstein-Hilbert action plus Gibbons-Hawking term is well-known:
\be
\label{eq:variationeh}
\delta I_{\text{gr}} = \int d^3 x (\text{eom}) + \frac{1}{16\pi G_N} \int d^2 x \sqrt{-\g}[\g^{ij} K - K^{ij}] \d \g_{ij},
\ee
and in Fefferman-Graham coordinates we find that:
\be
\label{eq:deltaigr}
\begin{split}
\delta I_\text{gr} &= \int d^3 x (\text{eom}) +  \frac{1}{16\pi G_N} \int d^2 x \sqrt{-g} \Big( \frac{1}{\r}g^{ij} + g'^{ij} - g^{ij} \tr(g^{-1}g')\Big) \d g_{ij},\\
\delta I_\text{ct} &= - \frac{1}{16\pi G_N} \frac{1}{\r}  \int d^2 x \, \sqrt{-g}g^{ij}\d g_{ij}.
\end{split}
\ee
As for the Chern-Simons part, we find
that
\be
\delta I_{\text{cs}} = \frac{1}{32 \pi G_N \m} \int d^3 x \, \sqrt{-G}\e^{\lambda\m\n}C_{\lambda \s}^\r R_{\n \m \r}^{\phantom{\n\m\r}\s} + \frac{1}{32 \pi G_N \m} \int d^2 x \, \sqrt{-\gamma} \e^{\lambda\m\n}n_\m \G_{\lambda\s}^\r  C_{\n\r}^\s,
\ee
with
\be
C_{\m\n}^\lambda = \d \G_{\m \n}^\lambda = \hf G^{\lambda \s} (\cdel_\m \d G_{\n\s} + \cdel_\n \d G_{\m \s} - \cdel_\s \d G_{\m \n})
\ee
and $n_\mu$ the outward pointing unit normal to the boundary and $\gamma_{ij}$ the induced metric on the boundary. Integrating the bulk part once more by parts, we find:
\begin{eqnarray}
\delta I_{\text{cs}} &=& - \frac{1}{32 \pi G_N \m} \int d^3 x \, \sqrt{-G}\e^{\lambda\m\n} (\cdel_\s R_{\n \m }^{\phantom{\n\m}\r\s}) \d G_{\lambda \r} \\
& & \qquad + \frac{1}{32 \pi G_N \m} \int d^2 x \, \sqrt{-\g} \e^{\lambda\m\n}( n_\m \G_{\lambda\s}^\r  C_{\n\r}^\s + n_\s  R_{\n \m}^{\phantom{\n\m}\r\s}\delta G_{\lambda \r} ) \nonumber
\end{eqnarray}
The first term eventually becomes the Cotton tensor in the equation of motion, using \eqref{eq:riemann3d} and the Bianchi identity.

Substituting now once more the Fefferman-Graham metric \eqref{eq:fgmetric}, we find $n_\m dx^\m = - d\r/(2\r)$ and the surface terms can be rewritten to yield:
\begin{align}
\delta I_{\text{cs}} &= \int d^3 x \, (\text{eom}) + \frac{1}{16 \pi G_N \m} \int d^2 x \sqrt{-g} \e^{ij}\Big(\hf \G_{ik}^l \delta \G_{jl}^k  + (g' g^{-1} \delta g)_{ij} - \r (g' g^{-1} \delta g')_{ij}
\nn \\ &\qquad \qquad + 2 \r (g'' g^{-1} \delta g)_{ij} - \r  (g'g^{-1}g'g^{-1}\d g)_{ij} \Big),\label{eq:deltaics}
\end{align}
with all covariant terms defined using $g_{ij}$. Notice that if $b_{(0)ij} = 0$ then all terms are finite in the limit where the radial cutoff $\r_0 \to 0$, in agreement with the above analysis for the on-shell action.

Combining then \eqref{eq:deltaigr} and \eqref{eq:deltaics}, the variation of the combined action $I_{\text{c}}$ defined in \eqref{eq:itot} is:
\begin{eqnarray}
\label{eq:variationaction}
\delta I_{\text{c}} &=& \frac{1}{16 \pi G_N} \int d^2 x \sqrt{-g} \Big\{ g'_{ij} - g_{ij}\tr(g^{-1}g')\Big\} (g^{-1}(\delta g) g^{-1})^{ij} \\
& & + \frac{1}{16 \pi G_N \m} \int d^2 x \sqrt{-g}  \Big\{ \hf A_{ij} - 2 \r \e_i^{\phantom{i}k}[g''_{kj} - \hf (g'g^{-1}g')_{kj}] - \e_i^{\phantom{i}k}g'_{kj} \Big\} (g^{-1}(\delta g) g^{-1})^{ij} \nonumber \\
&& + \frac{1}{16 \pi G_N \m}  \int d^2 x \sqrt{-g} \r\e_i^{\phantom{i}k}g'_{kj} (g^{-1}(\delta g') g^{-1})^{ij}. \nonumber
\end{eqnarray}
where the term $A_{ij}$ is a local term and is defined via:
\be
\label{eq:Xij}
\int d^2 x \sqrt{-g} \e^{ij}\G_{ik}^l \delta \G_{jl}^k = \int d^2 x \sqrt{-g}A^{ij} \delta g_{ij} \,.
\ee
Explicitly, we find:
\be
\begin{split}
A_{ij} &= \qt \Big[ \e^{kl}g^m_i g_{jn} + \e^{\phantom{i}l}_{i} g_j^m g_n^k - \e^{\phantom{j}l}_{j} g^{mk}g_{in} + (i \leftrightarrow j) \Big] \cdel_k \Gamma_{lm}^n\\
&= \Big[- \frac{1}{8}\e_i^{\phantom{i}k}\e_j^{\phantom{j}l} \e^{m n}  \cdel_l \del_m g_{nk} + (i\leftrightarrow j)\Big] + \frac{1}{4}\e^{kl}\cdel_k \del_l g_{ij}.
\end{split}
\ee

Notice that the last term in \eqref{eq:variationaction} involves $\delta g'_{ij}$ and therefore changes the variational principle for this action. Although one may explicitly check that it vanishes if $b_{(0)ij} = 0$ and for $\r_0 \to 0$ \cite{Kraus:2005zm}, this is no longer the case for nonzero $b_{(0)ij}$. As expected for a three-derivative bulk action, the on-shell action is a functional of both $g_{ij}$ and $g'_{ij}$ at the boundary and we can take functional derivatives with respect to both of them.

\subsection{One-point functions}
\label{sec:oneptfunctions}
From the previous section it follows that there are two independent sources that should be specified at the conformal boundary, which are asymptotically related to $g_{ij}$ and $g'_{ij}$. According to the asymptotic solution \eqref{eq:fgexpansion} obtained in section \ref{sec:asymptoticsoln} we can indeed independently specify both $b_{(0)ij}$ and $g_{(0)ij}$ and one can take these as the two boundary sources. These fields then source two operators which will be denoted $t_{ij}$ and $T_{ij}$, respectively, with $T_{ij}$ the usual energy-momentum tensor of the boundary theory. The standard AdS/CFT dictionary now dictates:
\be
\label{eq:defnTij}
\vev{T_{ij}} = \frac{- 4 \pi}{\sqrt{-g_{(0)}}} \frac{\d I}{\d g_{(0)}^{ij}}, \qquad \qquad \vev{t_{ij}} = \Big( \frac{- 4\pi}{\sqrt{-g_{(0)}}} \frac{\d I}{\d b_{(0)}^{ij}} \Big)_L ,
\ee
where the subscript `L' means a projection onto the chiral traceless component,
\be
(t_{ij})_{L} \equiv P_i^k(t_{kj} - \hf g_{kj} \tr(t)),
\ee
whose origin is explained in the next paragraph. The signs in \eqref{eq:defnTij} are explained in appendix \ref{app:analyticcont}. Notice that the on-shell action $I$ on the right-hand sides of \eqref{eq:defnTij} coincides with $I_{\text{c}}$ defined in \eqref{eq:itot} only to zeroth order in $b_{(0)ij}$, and as explained above additional boundary counterterms will be needed to render it finite to higher orders in $b_{(0)ij}$.

The projection onto the `L' component originates as follows.
Since $P_i^k b_{(0)kj} = \tr(b_{(0)}) = 0$,
$b_{(0)ij}$ has only a single nonvanishing
component. We can therefore only take functional derivatives with
respect to this component and we find that $t_{ij}$ only has one
component as well. For example, when we use lightcone coordinates and
the boundary metric is flat, $g_{(0)ij}dx^i dx^j = du dv$, then in our
conventions (see appendix \ref{app:analyticcont}) only $b_{(0)uu}$ is
nonzero. Correspondingly, the only non-zero component of
$t_{ij}$ is $t_{vv}$ and taking the
`L' piece projects onto this component.

To make contact with the regulated on-shell action which explicitly depends on $g_{ij}$ and $g'_{ij}$, we observe that:
\be
g_{(0)}^{ij} = \lim_{\r \to 0} (g^{ij} + \r \log(\r) g'^{ij}),\qquad \qquad b_{(0)ij} = \lim_{\r \to 0} \r g'^{ij},
\ee
and therefore the one-point functions can be obtained concretely by computing:
\be
\label{eq:vevsdefnsm1}
\begin{split}
\vev{t_{ij}} &= \lim_{\r \to 0}\Big( \frac{- 4 \pi}{\r \sqrt{-g}} \frac{\delta I}{\delta g'^{ij}} + \log(\r) \frac{4\pi}{\sqrt{-g}}\frac{\delta I}{\delta g^{ij}} \Big)_{L} ,\\
\vev{T_{ij}} &= \lim_{\r \to 0} \frac{- 4\pi}{\sqrt{-g}}\frac{\delta I}{\delta g^{ij}},
\end{split}
\ee
which are the main expressions that will be used in the following sections.

\subsubsection{Explicit expressions for vanishing $b_{(0)ij}$}
If we set $b_{(0)ij} = 0$ then the combined action $I_{\text{c}}$ is finite on-shell. Although we then cannot take functional derivatives with respect to $b_{(0)ij}$, we can still compute correlation functions involving the energy-momentum tensor by using the first equation in \eqref{eq:defnTij} with $I = I_{\text{c}}$. Explicitly, this means that we use \eqref{eq:variationaction} and substitute the expansion \eqref{eq:fgexpansion} with $b_{(0)ij} = 0$. This leads to the following one-point functions:
\begin{eqnarray}
\label{eq:tij}
\vev{T_{ij}} &\equiv & \lim_{\r \to 0} \frac{- 4\pi}{\sqrt{-g}}\frac{\delta I_{\text{c}}}{\delta g^{ij}}\\
&=&  \frac{1}{4 G_N}\Big( g'_{ij} - g_{ij}\tr(g^{-1}g') -  \frac{1}{\m}\Big( \hf \e_i^{\phantom{i}k} g'_{kj}  + \r \e_i^{\phantom{i}k} g''_{kj} + (i \leftrightarrow j) \Big) + \frac{1}{2\m} A_{ij}[g_{ij}] \Big) \nonumber \\
&=& \frac{1}{4 G_N}\Big(g_{(2)ij} + \hf R[g_{(0)}] g_{(0)ij} - \frac{1}{2\m}\Big(  \e_i^{\phantom{i}k} g_{(2)kj} + (i \leftrightarrow j)  \Big) - \frac{2}{\m}b_{(2)ij} + \frac{1}{2\m} A_{ij}[g_{(0)ij}] \Big) \nonumber
\end{eqnarray}
where we defined $\e_{i}^{\phantom{i}k}$ using $g_{(0)}$ and also used
the various properties of $b_{(2)ij}$ found above, in particular the
condition $\e_i^{\phantom{i}k} b_{(2)kj} = b_{(2)ij}$ which ensured
the absence of a logarithmic divergence. Notice that an extra sign
arises because we functionally differentiate with respect to the
inverse metric, whereas \eqref{eq:variationaction} uses a variation in
the metric itself. The expression with energy momentum tensor
with $b_{(0)ij} = b_{(2)ij} = 0$ was also derived previously
in \cite{Kraus:2005zm}. The authors of \cite{Grumiller:2008qz}
computed $T_{ij}$ for non-zero $b_{(2)ij}$ and flat $g_{(0)}$.
The result in equation (48) of \cite{Grumiller:2008qz} however is missing
the $b_{(2)}$ term.

%

Using $g_{(0)}$ to raise indices and define covariant derivatives and using the above properties of $b_{(2)ij}$ and $g_{(2)ij}$, we find the following Ward identities:
\be
\label{eq:wardids}
\begin{split}
\vev{T_i^i} &= \frac{1}{4 G_N}\Big( \hf R[g_{(0)}] + \frac{1}{2\m} A_i^i[g_{(0)}] \Big), \\
\cdel^j \vev{T_{ij}} &= \frac{1}{4 \m G_N}\Big(\qt \e_{ij}\cdel^j R[g_{(0)}] + \hf \cdel^j A_{ij}[g_{(0)}]\Big)\,.
\end{split}
\ee
These results agree with analogous computations in \cite{deHaro:2000xn,Solodukhin:2005ah}. We will discuss their interpretation in
the next section.

\paragraph{Example: conserved charges for the BTZ black hole\\}

The holographic energy momentum can be used to compute the conserved
charges, namely the mass and the angular momentum, for the
rotating BTZ black hole.
The metric can be written in Fefferman-Graham coordinates as:
\be
\begin{split}
ds^2 = \frac{d\r^2}{4\r^2} &- \Big[\frac{1}{\r} - \hf (r_+^2 + r_-^2) + \qt (r_+^2 - r_-^2)^2 \r \Big] dt^2 \\&+\Big[ \frac{1}{\r} + \hf (r_+^2 + r_-^2) + \qt (r_+^2 - r_-^2)^2 \r \Big] d\phi^2 + 2 r_+ r_- dt d\phi,
\end{split}
\ee
from which we find the following one-point function (using $\e_{t\phi} = -1$):
\be
\begin{split}
\vev{T_{tt}} &= \vev{T_{\phi \phi}} = \frac{1}{8 G_N}(r_+^2 + r_-^2 +  \frac{2}{\m} r_+ r_-), \\
\vev{T_{t\phi}} &= \frac{1}{8 G_N} (2 r_+ r_- + \frac{1}{\m} r_+^2 + r_-^2).
\end{split}
\ee
Notice that our normalization of the energy-momentum tensor differs by a factor of $2 \pi$ from that used in much of the AdS/CFT literature. We obtain the conserved charges:
\be
\begin{split}
M &= - \int d\phi T^t_t = \frac{\pi}{4 G_N}[r_+^2 + r_-^2 + \frac{2}{\m}r_+ r_-], \\
J &= - \int d\phi T^t_\phi = \frac{\pi}{4 G_N} [2 r_+ r_- + \frac{1}{\m}(r_+^2 + r_-^2)].
\end{split}
\ee
Up to the change in the overall normalization, these expressions agree with \cite{Moussa:2003fc,Kraus:2005zm} and in the Einstein case $\m \to \infty$ they reduce to the usual expressions. In lightcone coordinates $u = t + \phi, v = -t + \phi$ we find that
\be
\begin{split}
\vev{T_{uu}} &= \frac{1}{G_N} \Big( (1 + \frac{1}{\m}) (r_+^2 + r_-^2) + 2 ( \frac{1}{\m} + 1) r_+ r_- \Big),\\
\vev{T_{vv}} &= \frac{1}{G_N} \Big( (1 - \frac{1}{\m}) (r_+^2 + r_-^2) + 2 ( \frac{1}{\m} -1) r_+ r_- \Big).\\
\end{split}
\ee
so when $\m = 1$ only $T_{uu}$ is nonzero.

\section{Anomalies}
\label{sec:anomaly}

In this section we will discuss and interpret the anomalous Ward identities
\eqref{eq:wardids}. We will first consider the diffeomorphism anomaly and show that it agrees exactly with the expression expected from Wess-Zumino consistency conditions. We then discuss the Weyl anomaly and again find agreement with field theory expectations.

\subsection{Diffeomorphism anomaly}


The diffeomorphism Ward identity from  \eqref{eq:wardids} for $\m = 1$
reads
\be
\label{eq:diffanomaly}
\cdel^j \vev{T_{ij}} = \frac{1}{4 G_N} \Big( \frac{1}{4}\e_i^{\phantom{i}k}\cdel_k R[g_{(0)}] + \frac{1}{2} \cdel^j A_{ij}[g_{(0)}] \Big) \,.
\ee
The right-hand side is the diffeomorphism anomaly of the theory. A more explicit expression can be obtained following \cite{Bardeen:1984pm}. Consider a vector field $\z^i$. Then, under a diffeomorphism along $\z^i$ the metric change $\delta g_{ij} = \cdel_i \z_j + \cdel_j \z_i$ results in the following change in the connection coefficients:
\be
\delta \G_{ij}^k = \z^m\del_m \G_{ij}^k + (\del_i\z^m) \G_{mj}^k + (\del_j \z^m) \G_{im}^k - \G_{ij}^m \del_m \z^k + \del_i \del_j \z^k.
\ee
We may substitute this in \eqref{eq:Xij} and find that:
\be
\label{eq:cdelXij}
\begin{split}
&- 2 \int d^2 x \sqrt{-g}\z_j \cdel_i A^{ij} \\
&\qquad = \int d^2 x \sqrt{-g} \e^{ij}\G_{ik}^l \Big(\z^m\del_m \G_{jl}^k + (\del_j\z^m) \G_{ml}^k + (\del_l \z^m) \G_{jm}^k - \G_{jl}^m \del_m \z^k + \del_j \del_l \z^k \Big) \\
&\qquad =\int d^2 x \sqrt{-g} \Big( - \z^m \G_{mj}^i R \e_i^{\phantom{i}j} - (\del_j \z^i) R \e_i^{\phantom{i}j} - (\del_j \z^i)\e^{kl}\del_k \G_{li}^j\Big)\\
&\qquad = \int d^2 x \sqrt{-g} \z^i \Big( \e_i^{\phantom{i}j} \cdel_j R +  \e^{kl}\del_j \del_k \G_{li}^j\Big)
\end{split}
\ee
where the first term on the third line comes from the grouping the first two terms on the second line; to find it we used that $\e^{kl}\G_{ki}^j \G_{lj}^n \G_{mn}^i = 0$ in two dimensions. Substituting the explicit expression for $\cdel^i A_{ij}$ obtained from \eqref{eq:cdelXij} in \eqref{eq:diffanomaly} we obtain:
\be
\label{eq:consistentanomaly}
\cdel^j \vev{T_{ij}} = \frac{-1}{16 G_N} \e^{kl}\del_j \del_k \G_{li}^j.
\ee
As explained in \cite{Bardeen:1984pm,AlvarezGaume:1984dr}, this is precisely the two-dimensional diffeomorphism anomaly that satisfies the Wess-Zumino consistency conditions. In particular, in this case the consistency condition requires that the anomaly under a diffeomorphism along $\zeta$:
\be
H_{\zeta} = \int d^{2}x \sqrt{-g} \zeta^{i} \cdel^j \vev{T_{ij}},
\ee
satisfies
\be
E_{\zeta_1} H_{\zeta_2} - E_{\zeta_2} H_{\zeta_1} = H_{[\zeta_2,\zeta_1]},
\ee
where $E_{\zeta}$ denotes the action of a diffeomorphism with parameter $\zeta$.

The consistent anomaly \eqref{eq:consistentanomaly} is not covariant \cite{Bardeen:1984pm,AlvarezGaume:1984dr} and therefore $T_{ij}$ itself is not a covariant tensor either. One may try to remedy this by finding a symmetric local `improvement term' $Y_{ij}$ such that the new object $\hat T_{ij}$ defined as:
\be
\hat T_{ij} = T_{ij} + Y_{ij}
\ee
does transform as a tensor. This implies that $\cdel^i \hat T_{ij}$ is also covariant, resulting in a covariant diffeomorphism anomaly \cite{Bardeen:1984pm}. The covariant anomaly does not however satisfy the consistency conditions \cite{AlvarezGaume:1984dr} and therefore $\hat T_{ij}$ is not the variation of an effective action.

To better understand the form \eqref{eq:diffanomaly} of the diffeomorphism anomaly, we will now review the results summarized in \cite{Bardeen:1984pm}.\footnote{Our conventions differ as follows. Our  $T_{ij}$ has an extra $1/\sqrt{-g}$ as opposed to the analogous object in \cite{Bardeen:1984pm}; indeed, in our case $\hat T_{ij}$ is a tensor whereas in \cite{Bardeen:1984pm} it is a tensor density. The overall sign of the energy-momentum tensors is however the same. The connection $\G_{ij}^k$ in \cite{Bardeen:1984pm} is defined with an extra minus sign, but the Riemann curvature has the same sign. Finally, we always use a covariant $\e$-symbol whereas this is not the case in \cite{Bardeen:1984pm}.} As we will see shortly, one may obtain the covariant and the consistent anomaly as well as the improvement term starting from a single polynomial $P({\bf \Omega})$ of degree $d/2 + 1$ whose arguments are matrix-valued forms ${\bf \Omega}$. (In this section such forms are always written using bold face.) Although $P$ generally depends on the theory at hand, in $d=2$ we find that $P$ should be quadratic, leaving us with the unique possibility:
\be
P({\bf \Omega}) = a \Tr({\bf \Omega} \wedge {\bf \Omega}),
\ee
with a so far arbitrary normalization factor $a$. We will also write $P({\bf \Omega_1},{\bf \Omega_2}) = a \Tr({\bf \Omega_1} \wedge {\bf \Omega_2})$. Following the usual conventions \cite{Bardeen:1984pm, AlvarezGaume:1984dr}, we view the connection coefficients $\G_{ij}^k$ as matrix-valued one-forms,
\be
{\bf\G} \equiv \G_j^k = \G_{ij}^k dx^i,
\ee
and the Riemann tensor as a matrix-valued two-form,
\be
{\bf R} \equiv R_k^l = \hf R_{ijk}^{\phantom{ijk}l} dx^i \wedge dx^j.
\ee
The consistent anomaly can be found by solving a set of descent equations which follow from the consistency condition, see \cite{Bardeen:1984pm}. Using a matrix-valued zero-form ${\bf v} = v_i^j = \del_i \z^j$, the end result can be written as:
\be
H_{\zeta} \equiv \int d^2 x \sqrt {-g} \z_i \cdel_j T^{ij} = \int P(d {\bf v},{\bf \G}).
\ee
With the above form of $P$ this can be written more explicitly as:
\begin{multline}
\label{eq:consistentanomalybz}
\int d^2 x \sqrt {-g} \z_i \cdel_j T^{ij} = - a \int \Tr(d {\bf v} \wedge {\bf \G}) \\
= - a \int (\del_k \del_i \z^j) \G_{lj}^i dx^k \wedge dx^l = - a \int d^2 x \sqrt{-g} \e^{kl} (\del_k \del_i \z^j) \G_{lj}^i.
\end{multline}
Similarly, the covariant anomaly is obtained in \cite{Bardeen:1984pm} as:
\begin{multline}
\label{eq:covariantanomalybz}
\int d^2 x \z_i \cdel_j \hat T^{ij} = 2 \int P({\bf M},{\bf R}) = - a \int (\cdel_i \z^j) R_{klj}^{\phantom{klj}i} dx^k \wedge dx^l \\
= - a \int \sqrt{-g} (\cdel_i z^j) \e^{kl}R_{klj}^{\phantom{klj}i} = - a \int \sqrt{-g}(\cdel_i z^j) R \e_j^{\phantom{j}i}
\end{multline}
where ${\bf M} = - \cdel_i \z^j$ is again a matrix-valued $0$-form and $R$ is the usual Ricci scalar. Finally, the improvement term $Y_{ij}$ is given as:
\be
\int d^2 x \sqrt{-g} Y^{ij}\delta g_{ij} = 2 \int \Tr(\delta {\bf\G} \wedge {\bf X})
\ee
in terms of the variation of the connection and a matrix-valued one-form ${\bf X}$ given again in terms of $P$. We refer to \cite{Bardeen:1984pm} for the exact expression for ${\bf X}$, which for $d=2$ however reduces immediately to ${\bf X} = a {\bf \G}$. We therefore find:
\be
\label{eq:Yij}
\int d^2 x \sqrt{-g} Y^{ij}\delta g_{ij} = 2 a \int \sqrt{-g} \e^{ij}(\delta \G_{ik}^l) \G_{jl}^k.
\ee

Let us now compare these results with our holographically computed expressions. Comparing \eqref{eq:consistentanomaly} with \eqref{eq:consistentanomalybz} we find precise agreement provided that:
\be
a = \frac{1}{16 G_N}.
\ee
Furthermore, we are now able to understand our original expression \eqref{eq:diffanomaly}. Namely, it is exactly of the form:
\be
\label{eq:splitT}
\cdel^i T_{ij} = \cdel^i \hat T_{ij} - \cdel^i Y_{ij}.
\ee
To see this, observe that the first term on the right-hand side of \eqref{eq:diffanomaly} agrees precisely with \eqref{eq:covariantanomalybz} and the second term is precisely $1/(8G_N) \cdel^i A_{ij}$ as can be seen by comparing \eqref{eq:Yij} with \eqref{eq:Xij}. (This was recently noted in \cite{Hotta:2009zn} as well.)

Notice that the energy-momentum tensor postulated in \cite{Solodukhin:2005ah} does not include the term $\hf A_{ij}$ that we obtained in \eqref{eq:tij} from the variation of the on-shell supergravity action. The energy-momentum tensor of \cite{Solodukhin:2005ah} is therefore precisely the tensor $\hat T_{ij}$ defined above. In agreement with the above discussion, this $\hat T_{ij}$ is not obtained from an on-shell action and the anomaly found there is precisely the covariant anomaly \eqref{eq:covariantanomalybz}.

\subsection{Weyl anomaly}

For the Weyl anomaly we find from \eqref{eq:wardids}:
\be
\vev{T_i^i} = \frac{1}{8G_N} \Big( R[g_{(0)}] + A_i^i[g_{(0)}]\Big).
\ee
We have already discussed that the extra term $A_i^i[g_{(0)}]$ can be removed by hand. We then obtain the trace of the covariant energy-momentum tensor:
\be
\vev{\hat T_i^i} = \frac{1}{8G_N} R[g_{(0)}].
\ee
On the other hand, in our conventions we should have:
\be
\vev{\hat T_i^i} = \frac{c_L + c_R}{24} R[g_{(0)}]
\ee
and therefore:
\be
c_L + c_R = \frac{3}{G_N}
\ee
which agrees with the analysis in section \ref{sec:comparisonlcft} below.

\section{Linearized analysis}
\label{sec:linearizedanalysis}
In order to compute correlation functions involving the operator $t_{ij}$ as well, we will proceed perturbatively. In this section we therefore consider small perturbations $\delta G_{\m \n} = H_{\m \n}$ around the AdS$_3$ background. We will first linearize the bulk equations of motion and solve these asymptotically in order to isolate the divergent pieces in the combined action $I_{\text{c}}$ defined in \eqref{eq:itot}. We then renormalize this action to second order in the fluctuations. Taking functional derivatives as in \eqref{eq:vevsdefnsm1}, we obtain finite expressions for the one-point functions of $T_{ij}$ and $t_{ij}$ in terms of the subleading coefficients in the radial expansion of the perturbations. Afterwards, we find the full linearized bulk solutions for $H_{ij}$ so we can express these subleading pieces as nonlocal functionals of the sources $g_{(0)ij}$ and $b_{(0)ij}$. Finally, a second functional derivative then gives all boundary two-point functions involving $T_{ij}$ and $t_{ij}$. At the end of this section we compare our results with those expected from a logarithmic CFT (LCFT) and find complete agreement.

\subsection{Linearized equations of motion}
We will now linearize the equations of section \ref{sec:fgeom} around an empty AdS background solution. We work in Poincar\'e coordinates where the background metric $G_{\m \n}$ has the form
\be
\label{eq:emptyadspoincare}
G_{\m \n}dx^\m dx^\n = \frac{d\r^2}{4\r^2} + \frac{1}{\r} \eta_{ij}dx^i dx^j.
\ee
An earlier investigation of the linearized equations around this background can be found in \cite{Carlip:2008jk,Carlip:2008eq}. As we work in Fefferman-Graham coordinates, it is natural to pick a radial-axial gauge for the fluctuations as well. Thus we set $H_{\r\r} = H_{\r i} = 0$ and define $h_{ij} \equiv \delta g_{ij} = H_{ij}/\r$. We therefore substitute
\be
g_{ij} = \eta_{ij} + h_{ij}
\ee
into the equations of motion \eqref{eq:eomfgform}. To leading order in $h_{ij}$ we find:
\be
\label{eq:linearizedeom}
\begin{split}
&- \tr(h'') + \frac{1}{2\m} \e^{ij} \del_i \del^m h'_{mj} = 0,\\
& 2 \r\del^k h''_{ik} + \del^k h'_{ik} + \m \e^{jk}\del_k h'_{ij} - \del_i \tr(h') = 0,\\
& - h''_{ij} + \eta_{ij}\hf \tr(h'')  \\
&\qquad + \frac{1}{\m}\e_i^{\phantom{i}k} \Big[ \qt \del_k \del^l h'_{lj} + \qt \del_j \del^l h'_{lk} - \hf \del_k \del_j \tr(h') + 2 \r h'''_{jk}
+3 h''_{jk} \Big] + (i \leftrightarrow j) = 0,
\end{split}
\ee
and for the trace equation $R = -6$ we obtain:
\be
\label{eq:linearizedR}
- 4 \r \tr(h'') + \tilde R(h) + 2 \tr(h') = 0,
\ee
with $\tilde R[h]$ the linearized curvature of $\eta_{ij} + h_{ij}$, which can be explicitly written as
\be
\tilde R[h] = \cdel^i \cdel^j h_{ij} - \cdel^i \cdel_i \tr(h)\,.
\ee
Notice that all covariant symbols and traces in the above equations are defined using the background metric $\eta_{ij}$.

We also obtained the linearized equations of motion in global
coordinates, which can be found in appendix \ref{sec:applinearizedeom}.
The analysis in global coordinates would be useful should one want to
compute directly\footnote{Alternatively, one can obtain the correlators
on $R \times S^1$ from the ones
on $R^2$ by using the fact that $R \times S^1$ is conformally related
to $R^2$ and finite Weyl transformations in the boundary theory
can be implemented
by specific bulk diffeomorphisms \cite{Skenderis:2000in} (whose infinitesimal
form was derived first in \cite{Imbimbo:1999bj}).}
the correlators of the CFT on $R \times S^1$ rather than $R^2$.

\subsection{Holographic renormalization}
In this subsection we consider the holographic renormalization of the on-shell action. Since we work at the linearized level, we compute the on-shell action to second order in the perturbations around the Poincar\'e background. We isolate the divergences to that order and compute the necessary covariant counterterms to cancel these divergences.

\subsubsection{Asymptotic analysis}
\label{sec:linearizedeom}
We begin by substituting the asymptotic expansion for $h_{ij}$:
\be
\label{eq:expansionh}
h_{ij} = b_{(0)ij} \log(\r) + h_{(0)ij} + b_{(2)ij} \r \log(\r) + h_{(2)ij}\r + \ldots
\ee
into the linearized equations of motion \eqref{eq:linearizedeom} and \eqref{eq:linearizedR}. We find from the linearization of the asymptotic analysis above that:
\be
\label{eq:linearizedeompoincare}
\begin{split}
\tr(b_{(0)}) &= 0,\\
b_{ij} + \e_i^{\phantom{i}k} b_{kj} &= 0,\\
\tr(b_{(2)}) &= - \hf \tilde R[b_{(0)}] = -\hf \del^i \del^j b_{(0)ij},\\
\tr(h_{(2)}) &= - \hf \tilde R[h_{(0)}] + \tr(b_{(2)}) ,\\
b_{(2)ij} - \e_i^{\phantom{i}k} b_{(2)kj} &= \hf \eta_{ij}\tr(b_{(2)}) + \qt \e_i^{\phantom{i}k} (\del_k \del^l b_{(0)lj} + \del_j \del^l b_{(0)lk}),\\
\del^j \Big( b_{(2)ij} - 3 \e_i^{\phantom{i}k}b_{(2)kj} & + 2 \bar P_i^k h_{(2)kj} - 2 \bar P_i^k \eta_{kj}(\tr(h_{(2)}) + \tr(b_{(2)}) ) \Big) = 0,
\end{split}
\ee
where all covariant symbols and traces are defined using $\eta_{ij}$ and $\tilde R[h]$ again denotes the linearized curvature of the metric $\eta_{ij} + h_{ij}$.

\subsubsection{On-shell action and counterterms}
\label{sec:linearizedactions}
The next step is to substitute the asymptotic expansion \eqref{eq:expansionh}, together with the constraints \eqref{eq:linearizedeompoincare}, into the on-shell action \eqref{eq:itot}. We then isolate the divergences and find the necessary counterterm action that makes the action finite to second order $h_{ij}$.

Expanding the on-shell action \eqref{eq:itot} in $h_{ij}$, we find that the first-order term vanishes, since it gives a term proportional to the bulk equations of motion plus the surface terms of \eqref{eq:variationaction}, which vanish identically for the Poincar\'e background. At the second order we find:
\be
\label{eq:variationactionpoincare}
I_2 = \frac{1}{32 \pi G_N} \int d^2 x \Big(h'_{ij} - \eta_{ij}\tr(h') - 2 \r \e_i^{\phantom{i}k}h''_{kj} - \e_i^{\phantom{i}k} h'_{kj}\Big) h^{ij}.
\ee
Notice that there are no contributions from the $A_{ij}$-term for the Poincar\'e background, as can be seen easily from its definition \eqref{eq:Xij}. If we now substitute the expansion \eqref{eq:expansionh} and use the linearized equations of motion \eqref{eq:linearizedeompoincare} then we find a logarithmic divergence of the form:
\be
I_{2} = \frac{1}{32\pi G_N}\int d^2 x \Big(\hf \tr(h_{(0)})\tilde R[b_{(0)}] - 2 b_{(2)ij}b^{ij}_{(0)} - \hf h_{(0)i}^k \del^i \del^j b_{(0)jk} \Big) \log(\r) + \ldots
\ee
The next step in the holographic renormalization is to invert the series and rewrite the divergent terms in terms of $h_{ij}$ plus finite corrections. This gives:
\be
\begin{split}
\log(\r) b_{(0)ij} &= h_{ij} + \ldots,\\
h_{(0)ij} &= h_{ij} - \r \log(\r) h'_{ij} + \ldots,\\
\log(\r) b_{(2)ij}b^{ij}_{(0)} &= \hf \r h'_{ij}h'^{ij} + \ldots,\\
\end{split}
\ee
and we also have:
\be
\tr(h_{(0)}) \tilde R[b_{(0)}] = 2 h_{(0)i}^k \del^i \del^j b_{(0)jk} - h^{ij}\del^k \del_k b_{(0)ij},
\ee
from which we find that this divergence is cancelled by adding the following counterterm action:
\be
\label{eq:i2ct}
I_{2,\text{ct}} = \frac{1}{32\pi G_N} \int d^2 x \Big( \qt h^{ij}\del^k \del_k h_{ij}  + \r h'_{ij} h'^{ij} - \frac{1}{4} h_i^j \del^i \del^k h_{kj} \Big).
\ee
This action can be written in a covariant form as follows. The background induced metric is written $\gamma_{ij} = \eta_{ij}/\r$ and its deviation $h_{ij}/\r = \s_{ij}$. The extrinsic curvature $K_i^j = - \d_i^j + \r g'^j_i$ and its deviation is $\tilde K_i^j[h] = \r h'^j_i$. In this notation, the counterterm action becomes:
\be
I_{2,\text{ct}} = \frac{1}{32\pi G_N} \int d^2 x \sqrt{-\g} \Big(\qt \s^{ij} \cdel^k \cdel_k \s_{ij} + \tilde K_{ij}[h] \tilde K^{ij}[h] - \frac{1}{4} \s_i^j \cdel^i \cdel^k \s_{kj} \Big),
\ee
where indices are now raised and covariant derivatives and traces are defined using $\g_{ij}$.

Notice that the counterterm action involves the extrinsic curvature $K_{ij}$ as well. Such a term would not be allowed in pure Einstein theory as it would lead to an incorrect variational principle. On the other hand, for TMG we already found that the variational principle is different. In particular, the higher-derivative terms allow for the specification of both $\g_{ij}$ and $K_{ij}$ at the boundary and therefore we are also allowed to use $K_{ij}$ in the boundary counterterm action.

\subsubsection{One-point functions}
For the total action at this order $I_{2,\text{tot}} = I_{2} + I_{2,\text{ct}}$ we find the variations:
\begin{eqnarray}
\label{eq:variationpoinc}
\frac{\delta I_{2,\text{tot}}}{\delta h^{ij}} &=& \frac{1}{16\pi G_N} \Big(h'_{ij} - \eta_{ij}\tr(h') - 2 \r \e_i^{\phantom{i}k}h''_{kj} - \e_i^{\phantom{i}k} h'_{kj} + \hf \tilde A_{ij}[h] + \frac{1}{4} \del^k \del_k h_{ij} - \frac{1}{4}\del_i \del^k h_{kj}\Big),\nonumber \\
\frac{\delta I_{2,\text{tot}}}{\delta h'^{ij}} &=& \frac{1}{16 \pi G_N} \r (\delta_i^k + \e_i^{\phantom{i}k})h'_{kj},
\end{eqnarray}
with $\tilde A_{ij}[h]$ the linearization of $A_{ij}$ as defined in \eqref{eq:Xij}:
\be
\tilde A_{ij}[h] = \frac{1}{4}\e_i^{\phantom{i}k} (\del_j \del^l h_{kl} - \del^l \del_l h_{kj}) + (i \leftrightarrow j).
\ee
We now substitute the expansion \eqref{eq:expansionh} and find:
\be
\begin{split}
\frac{\delta I_{2,\text{tot}}}{\delta h^{ij}} &= \frac{1}{16 \pi G_N}\Big\{ b_{(2)ij} - 3 \e_i^{\phantom{i}k} b_{(2)kj} + 2 \bar P_i^k h_{(2)kj} + \eta_{ij}\Big(\hf \tilde R[h_{(0)}] + \tilde R[b_{(0)}]\Big) \\ & \qquad + \hf \bar P_i^k\Big(\del^l \del_l h_{(0)kj} - \del_j \del^l h_{(0)lk}\Big) \Big\},\\
\frac{\delta I_{2,\text{tot}}}{\delta h'^{ij}} &= \frac{\r}{8 \pi G_N} P_i^k \Big(b_{(2)kj} \log(\r) + b_{(2)kj} + h_{(2)kj}\Big),
\end{split}
\ee
where we dropped terms that vanish as $\r \to 0$ and do not contribute below.
In the above formulas symmetrization in $i$ and $j$ is implicit.
When $b_{(0)ij} = 0$ we can compare the first of these expressions with \eqref{eq:tij} and we find that the additional counterterms only change the local terms.

Using \eqref{eq:vevsdefnsm1} and taking into account an extra sign from the fact that $g^{ij} = \eta^{ij} - h^{ij}$, we obtain the following explicit expression for the one-point functions:
\be
\label{eq:linearizedoneptfns}
\begin{split}
\vev{T_{ij}} &= \lim_{\r \to 0} \frac{4\pi}{\sqrt{- \eta}} \frac{\delta I_{2,\text{tot}}}{\delta h^{ij}} \\
&= \frac{1}{4 G_N} \Big\{ b_{(2)ij} - 3 \e_i^{\phantom{i}k} b_{(2)kj} + 2 \bar P_i^k h_{(2)kj} + \eta_{ij}\Big(\hf \tilde R[h_{(0)}] + \tilde R[b_{(0)}]\Big) \\ & \qquad + \hf \bar P_i^k\Big(\del^l \del_l h_{(0)kj} - \del_j \del^l h_{(0)lk}\Big) \Big\},\\
\vev{t_{ij}} &= \lim_{\r \to 0} \Big( \frac{- 4 \pi}{\r \sqrt{-g}} \frac{\delta I}{\delta h'^{ij}} - \log(\r) \frac{4\pi}{\sqrt{-\eta}}\frac{\delta I}{\delta h^{ij}} \Big)_{L} = \frac{1}{2 G_N}\Big(b_{(2)ij} + h_{(2)ij} \Big)_{L},\\
\end{split}
\ee
where we note that the projection to the $L$-component in $\vev{t_{ij}}$
also removes (divergent) terms of the form
$\eta_{ij}(\ldots)$ or $\bar P_i^k(\ldots)_{kj}$.
.

\subsection{Exact solutions}
\label{sec:exactsolutions}
In this subsection we solve the linearized equations of motion given in section \ref{sec:linearizedeom}. From the explicit solutions we find below, we can obtain the subleading terms $b_{(2)ij}$ and $h_{(2)ij}$ that enter in \eqref{eq:linearizedoneptfns} as nonlocal functionals of $g_{(0)ij}$ and $b_{(0)ij}$. This will allow us to carry out the second functional differentiation required to obtain the two-point functions.

In explicitly solving the fluctuation equations it is convenient to
Wick rotate and
work in Euclidean signature; the procedure for analytic continuation is
explained in detail in appendix \ref{app:analyticcont}. Concretely,
one starts from the metric \eqref{eq:emptyadspoincare},
introduces lightcone coordinates $u = t+x$, $v = -t + x$, and replaces
$v \to z$, $u \to \bar z$ with $(z,\bar z)$ complex boundary
coordinates. The background metric then has the form:
\be
ds^2 = \frac{d\r^2}{4\r^2} + \frac{1}{\r}dz d\bar z.
\ee
We will employ the notation $\del \equiv \del_z$ and $\bar \del \equiv \del_{\bar z}$ below.

In these coordinates, the linearized equations of motion \eqref{eq:linearizedeom} and \eqref{eq:linearizedR} become:
\be
\label{eq:lightconepoinc}
\begin{split}
-\bar \del (1+\m) h_{z\bar z}' + \del(1 + \m) h_{\bar z\bar z}'+2 \r \left(\del h_{\bar z\bar z}''+\bar \del
h_{z\bar z}''\right) = 0\\
\del (1-\m) h_{z\bar z}'-\bar \del (1-\m) h_{zz}'-2 \r \left(\del h_{z\bar z}''+\bar \del h_{zz}''\right) = 0\\
-\bar \del^2 h_{z\bar z}' +\bar \del \del h_{\bar z\bar z}'+(3 + \m) h_{\bar z\bar z}''+ 2 \r h_{\bar z\bar z}^{(3)} = 0\\
-\del^2 h_{z\bar z}'+\bar \del \del h_{zz}'+(3 -\m) h_{zz}''+2
\r h_{zz}^{(3)} = 0\\
\del^2 h_{\bar z\bar z}'-\bar \del^2 h_{zz}'+ 2 \m h_{z\bar z}'' = 0\\
\del^2 h_{\bar z\bar z}-2 \bar \del \del h_{z\bar z}+\bar \del^2 h_{zz}+ 2 h_{z\bar z}'- 4 \r h_{z\bar z}''= 0,
\end{split}
\ee
where again we have temporarily reinstated $\m$ for later use. From these equations it is straightforward to verify that $h_{z \bar z}''$ satisfies a Bessel-like equation:
\be
4 \r^2 h_{z\bar z}^{(4)} + 8 \r h_{z\bar z}^{(3)} + (4 \r \bar \del \del - \m^2 + 1)h_{z\bar z}'' = 0,
\ee
which has the general solution:
\be
\label{eq:Tbessel}
h_{z\bar z}'' = \r^{-1/2} K_{\m} (q \sqrt{\r}) \alpha + \r^{-1/2} I_{\m} (q \sqrt{\r}) \beta,
\ee
with $\a$ and $\b$ arbitrary functions of $u$ and $v$ and we defined $q = \sqrt{- 4 \bar \del \del}$. Passing to momentum space, we have $q \geq 0$ and only $K_{\m}$ is regular as $\r \to \infty$ and we therefore set $\b = 0$.

As a sidenote, in real time it is possible that $q < 0$ and then both
solutions have a power-law divergence as $\r \to \infty$. A solution
that is regular at $\r \to \infty$ can nevertheless be constructed
from them using an infinite number of these modes
\cite{Carlip:2008jk,Carlip:2008eq}; see also \cite{Skenderis:2008dg}
for an explicit example. Alternatively, one can solve the fluctuation
equation using global coordinates. In any case,
since we work in Euclidean signature such singular
behavior for the individual modes is absent and there is no need to
worry about these issues.


We can integrate \eqref{eq:Tbessel} twice to find an explicit solution
for $h_{z \bar z}$ which for general $\m$ involves an integral of the
hypergeometric functions ${}_1F_2$. Notice also that as  $\m \to \infty$
the linearized Einstein equations become $h_{z \bar z}'' = 0$, so the
radial dependence of the perturbation is linear in $\r$. This correctly
reproduces the linearization of the exact solution of the non-linear
vacuum Einstein equation in three dimension
in Fefferman-Graham coordinates given in \cite{Skenderis:1999nb},
which has a Fefferman-Graham expansion that terminates at $\r^2$.

For the other components, the last two equations in \eqref{eq:lightconepoinc} may be exploited to find that:
\be
\label{eq:UpVp}
\begin{split}
2 \del^2 h_{\bar z\bar z}' = 4 \r h_{z\bar z}^{(3)} + 2 (1-\m) h_{z\bar z}'' + 2 \bar \del \del h_{z\bar z}',\\
2 \bar \del^2 h_{zz}' = 4 \r h_{z\bar z}^{(3)} + 2 (1+\m) h_{z\bar z}'' + 2 \bar \del \del h_{z\bar z}',
\end{split}
\ee
which allows us to completely solve the system.

\subsubsection{Solutions for $\m = 1$}
In contrast to the case for general $\m$, for $\m = 1$ one may use the modified Bessel equation:
\be
\del_x^2 \Big(\sqrt x K_1 (\sqrt x)\Big) = \frac{1}{4 \sqrt x} K_1(\sqrt x)
\ee
to integrate \eqref{eq:Tbessel} twice giving:
\be
h_{z\bar z} = B_{z\bar z} \del^2 c_{0} + c_{1} \r + c_{2},
\ee
where $c_i$ are integration constants which are arbitrary functions of $\bar z$ and $z$ and we defined
\be
B_{z\bar z} \equiv  - \frac{2 \sqrt{\r}}{q} K_1(q \sqrt{\r}). 
\ee
Notice that it is convenient to express $h_{z \bar z}''$ as:
\be
h_{z\bar z}'' =  - \frac{1}{\r} B_{z\bar z} \bar \del \del^3 c_{0}.
\ee
Integrating \eqref{eq:UpVp} then results in:
\be
\label{eq:UpVpintegrated}
\begin{split}
h_{\bar z\bar z} &= - B_{z\bar z} \del \bar \del c_{0} - 2 B_{z\bar z}' c_{0} + \frac{\bar \del}{\del} c_{1} \r + c_3,\\
h_{zz} &= - B_{z\bar z} \frac{\del^3}{\bar \del} c_{0} + \frac{\del}{\bar \del} c_{1} \r + c_4,
\end{split}
\ee
and the last equation in \eqref{eq:lightconepoinc} gives the constraint:
\be
\label{eq:constraintconstants}
2 c_1 + \bar \del^2 c_4 + \del^2 c_3 - 2 \bar \del \del c_2 = 0,
\ee
i.e. $c_1$ is not an independent integration constant, but is determined in terms of the other integration constants.

Near the boundary $\r \to 0$ we have the following expansion:
\be
B_{z\bar z} = - \frac{2}{q^2} - \frac{\r}{2}(2\g -1) - \r \log(\frac{q\sqrt\r}{2}) - \frac{q^2\r^2}{8} \log (\frac{q\sqrt\r}{2}) + \ldots,
\ee
with $\g$ the Euler-Mascheroni constant. Substitution in \eqref{eq:UpVpintegrated} then yields the expansions for the components:
\begin{align}
\label{eq:radialexpcomponents}
h_{z\bar z} &= h_{(0) z\bar z} - \frac{1}{2} \r \log(\r) \del^2 b_{(0)\bar z \bar z} + \r h_{(2) z\bar z} + \ldots, \\
h_{\bar z\bar z} &= b_{(0)\bar z \bar z}\log(\r) + h_{(0) \bar z\bar z} - \frac{1}{2} \r \log(\r)  \bar \del \del b_{(0)\bar z \bar z} + \r \Big[ \frac{\bar \del}{\del} h_{(2) z\bar z} + \frac{4 \g - 3}{2}\bar \del \del b_{(0)\bar z \bar z}\Big] + \ldots, \nn \\
h_{zz} &= h_{(0) zz} + \frac{1}{2}\r \log(\r) \frac{\del^3}{\bar \del} b_{(0)\bar z \bar z} + \r \Big[ \Big( 2 \g -1 + 2 \log(\frac{q}{2}) \Big)\frac{\del^3}{\bar \del} b_{(0)\bar z \bar z} + \frac{\del}{\bar \del}h_{(2) z\bar z}\Big] + \ldots,\nn
\end{align}
where the boundary sources $h_{(0)ij}$ and $b_{(0)\bar z \bar z}$ are given by the following combinations of the integration constants $c_i$:
\begin{align}
h_{(0) z\bar z} &= c_2 - \frac{2}{q^2}\del^2 c_0 & h_{(0) zz} &= c_4 - \hf \frac{\del^2}{\bar \del^2} c_0 \nn \\
h_{(0) \bar z\bar z} &= c_3 - \hf c_0 + 2 \g c_0 + 2 \log(\frac{q}{2}) c_0 & b_{(0)\bar z \bar z} &= c_0.
\end{align}
The normalizable mode is the combination:
\be
h_{(2) z\bar z} = c_1 - \frac{2\g -1}{2}\del^2 c_0 - \log(\frac{q}{2}) \del^2 c_0,
\ee
which using \eqref{eq:constraintconstants} is determined by the boundary sources via:
\be
\label{eq:linearizedtrace}
h_{(2) z\bar z}  = -\frac{1}{2}\del^2 h_{(0) \bar z\bar z} - \frac{1}{2}\bar \del^2 h_{(0) zz} + \bar \del \del h_{(0) z\bar z} - \frac{1}{2} \del^2 b_{(0)\bar z \bar z}.
\ee
This is indeed the linearized form of \eqref{eq:trg2} and \eqref{eq:trh} combined.
Notice also that the radial expansion \eqref{eq:radialexpcomponents} indeed shows the same asymptotic behavior as \eqref{eq:fgexpansion} in section \ref{sec:asymptoticsoln}.

\subsection{Two-point functions}
\label{sec:twoptfunctions}
Substituting the solutions that we found above into the holographic one point functions \eqref{eq:linearizedoneptfns}, we find that:
\be
\label{eq:exactoneptfn}
\begin{split}
\vev{t_{zz}} &= \frac{-1}{4 G_N}\Big( (4\g - 1) \frac{\del^3}{\bar \del}b_{(0)\bar z \bar z} + 4 \log(\frac{q}{2}) \frac{\del^3}{\bar \del} b_{(0)\bar z \bar z} + 2 \frac{\del}{\bar \del}h_{(2) z\bar z} \Big), \\
\vev{T_{z \bar z}} &= \text{local},\\
\vev{T_{zz}} &= - \frac{1}{4 G_N}\Big( \frac{\del^3}{\bar \del} b_{(0)zz} + \text{local}\Big),\\
\vev{T_{\bar z \bar z}} &= \frac{1}{2 G_N} \Big( \frac{\bar \del}{\del}h_{(2)z \bar z} + \text{local}\Big),
\end{split}
\ee
where the local pieces correspond to finite contact terms.

We now turn to the position space expressions for the two-point functions. These are obtained via the following functional differentiations:
\be
\vev{T_{ij}\ldots} = i \frac{4 \pi}{\sqrt{- g_{(0)}}} \frac{\delta}{\delta g^{ij}_{(0)}} \vev{\ldots}, \qquad \qquad \vev{t_{ij}\ldots} = i \frac{4 \pi}{\sqrt{- g_{(0)}}} \frac{\delta}{\delta b^{ij}_{(0)}} \vev{\ldots},
\ee
where the prefactors are explained in appendix \ref{app:analyticcont}. Notice that in complex coordinates $ds^2 = dz d\bar z$ so $\sqrt{- g_{(0)}} = 1/2$ whilst
in our case $g^{ij} = \eta^{ij} - h^{ij}$ and therefore
\be
\frac{\d}{\d g^{ij}} = - \frac{\d}{\d h^{ij}} = - \eta_{ik} \eta_{jl} \frac{\d}{\d h_{kl}}
\ee
which in complex coordinates becomes:
\be
\frac{\d}{\d g^{zz}_{(0)}} = - \qt \frac{\d}{\d h_{\bar z \bar z}}, \qquad \qquad \frac{\d}{\d g^{\bar z \bar z}_{(0)}} = - \qt \frac{\d}{\d h_{z z}}.
\ee
Functionally differentiating the one point functions thus results in:
\be
\begin{split}
\vev{t_{zz}(z,\bar z)t_{zz}(0)} &= - \frac{2 \pi i}{G_N} \Big[ (\g - \qt)  + \log(\frac{q}{2}) \Big] \frac{\del^3}{\bar \del} \delta^2 (z,\bar z)\\
\vev{t_{zz}(z,\bar z)T_{zz}(0)} &= - \frac{i \pi}{2 G_N} \frac{\del^3}{\bar \del} \delta^2 (z,\bar z)\\
\vev{T_{\bar z \bar z}(z,\bar z)T_{\bar z \bar z}(0)} &= \frac{i \pi}{2 G_N} \frac{\bar \del^3}{\del} \delta^2 (z,\bar z)
\end{split}
\ee
whilst $\vev{t_{zz} T_{\bar z \bar z}} = \vev{T_{\bar z \bar z}T_{zz}}  = \vev{T_{zz}T_{zz}} = 0$ up to contact terms.

These expressions can be evaluated using the following set of identities. First notice that:
\be
- 2i \d^2(z,\bar z) = \d(x) \d(\tau), \qquad \qquad 4 \del \bar \del = \del_\tau^2 + \del_x^2.
\ee
The former of these is obtained by requesting $\int d^2 z \delta^2(z,\bar z) = 1$ and $\hf \int d^2 z(\ldots)  = - i \int d^2 x (\ldots)$. We also need the following integral, which can be directly computed using the properties of the Bessel function $J_0(x)$:
\be
\label{eq:fourierqa}
\frac{1}{4\pi^2} \int d\omega dk e^{i\omega \tau + i k x} \frac{1}{(\omega^2 + k^2)^{\a/2}} = \frac{1}{\pi} 2^{-\a} \frac{\G(1-\a/2)}{\G(\a/2)}(\tau^2 + x^2)^{-1 + (\a/2)}.
\ee
Taking the limit $\a = 2$ on both sides gives the identity:
\be
\frac{1}{\del \bar \del} \d^2(z,\bar z) = \frac{2i}{\del_\tau^2 + \del_x^2}\d^2(x,y) = \frac{i}{2\pi} \log(m^2(\tau^2 + x^2)) = \frac{i}{2\pi} \log(m^2 |z|^2)
\ee
where $m$ is a scale. By differentiating both sides in \eqref{eq:fourierqa} with respect to $\a$ we also find:
\be
\log(q) \frac{1}{\del \bar \del} \d^2(z,\bar z) = \log(q) \frac{2i}{\del_\tau^2 + \del_x^2} \d^2(x,y) = - \frac{i}{8 \pi} \log^2(m^2(\tau^2 + x^2)) = - \frac{i}{8\pi} \log^2(m^2|z|^2).
\ee
Using these expressions the two-point functions become:
\be \label{two-pt}
\begin{split}
\vev{t_{zz}(z,\bar z)t_{zz}(0)} &= \frac{1}{4 G_N} \del^4 [ B_m \log(m^2 |z|^2) - \log^2( m^2 |z|^2) ] \\
&= \frac{1}{2 G_N} \frac{- 3 B_m  - 11 + 6 \log(m^2 |z|^2)}{z^4},\\
\vev{t_{zz}(z,\bar z)T_{zz}(0)} &= \frac{1}{4 G_N} \del^4 \log(m^2 |z|^2) = \frac{- 3/(2 G_N)}{z^4},\\
\vev{T_{\bar z \bar z}(z,\bar z)T_{\bar z \bar z}(0)} &= \frac{3/(2 G_N)}{\bar z^4},\\
\end{split}
\ee
where $B_m$ is a scale-dependent constant that can be changed by rescaling $m$ in the first line. In fact, the entire non-logarithmic piece in the second line can also be removed from the correlation function by redefining $t \to t - (3 B_m + 11) T_{zz}/6$. This transformation is familiar from logarithmic CFT as we review in appendix \ref{sec:applcft}.

\subsubsection{Comparison to logarithmic CFT}
\label{sec:comparisonlcft}
The expressions above agree with general expectations from a logarithmic CFT, see appendix \ref{sec:applcft} for an introduction. The central charges can be computed as follows. From the two-point functions of $T_{\bar z \bar z}$ and $T_{zz}$, which should be of the form:
\be
\label{eq:clcr}
\vev{T_{zz}T_{zz}} = \frac{c_L}{2 z^4}, \qquad \qquad   \vev{T_{\bar z \bar z}T_{\bar z \bar z}} = \frac{c_R}{2 \bar z^4},
\ee
we find that
\be
c_L = 0, \qquad \qquad c_R =  \frac{3}{G_N},
\ee
which agrees with \cite{Li:2008dq}. As discussed in appendix \ref{sec:applcft}
two point functions of a logarithmic pair of operators $(T,t)$ in a LCFT have the structure:
\begin{eqnarray}
\< T(z) T(0) \> &=& 0; \qquad \< T(z) t (0,0) \> = \frac{b}{2z^4}; \label{eq:bdef} \\
\< t(z, \bar{z}) t(0,0) \> &=& \frac{- b \log (m^2 |z|^2)}{z^4}. \nonumber
\end{eqnarray}
Note that by rescaling the operator $t$ the coefficients of the non-zero two point functions can be changed; there is however a distinguished
normalization of the operator in which the two point functions take this form, and the coefficient $b$ is sometimes referred to as the new
anomaly, see \cite{Gurarie:2004ce}.
Comparing these expressions with (\ref{two-pt}) we see that our holographic two point functions indeed have the structure expected from a LCFT and
the coefficient $b$ is:
\be
b = - \frac{3}{G_N}.
\ee
This value will be confirmed below in the analysis for general $\mu$.


\section{Linearized analysis for general $\mu$}
\label{sec:linearizedgeneralmu}
In this section we repeat the linearized analysis of section \ref{sec:linearizedanalysis} for general $\mu$ around the Poincar\'e background. We define:
\be
\lambda = \hf(\m - 1), \qquad \qquad \m = 2 \lambda + 1,
\ee
and we work around $\lambda = 0$.

\subsection{Asymptotic analysis}
The linearized equations of motion give the most general asymptotic form of the solution:
\be
\label{eq:fgexpansiongeneralmu}
h_{ij} = h_{(-2\lambda)ij}\r^{-\lambda} + h_{(0)ij} + h_{(2)ij} \r  + h_{(2-2\lambda)ij}\r^{1-\lambda} + h_{(2 + 2\lambda)ij} \r^{\lambda + 1} + \ldots,
\ee
with the conditions:
\begin{align}
\tr(h_{(-2\lambda)}) &= 0 & P_i^k h_{(-2\lambda)kj} &= 0 & \tr(h_{(2)}) &= - \hf \tilde R [h_{(0)}] \nn\\
\tr(h_{(2-2\lambda)}) &= \frac{-\tilde R [h_{(-2\lambda)}]}{2(1-\lambda)(2\lambda + 1)}  & \tr(h_{(2\lambda + 2)}) &= 0 & \bar P_i^k h_{(2\lambda + 2)kj} &= 0
\end{align}
\[
\begin{split}
h_{(2-2\lambda)ij} + \frac{2 \lambda -1}{2\lambda + 1}\e_i^{\phantom{i}k}h_{(2-2\lambda)kj} &= \hf \eta_{ij}\tr(h_{(2-2\lambda)}) + \frac{\e_i^{\phantom{i}k} (\del_k \del^l h_{(-2\lambda)lj} + \del_j \del^l h_{(-2\lambda)lk})}{4 (1-\lambda)(2\lambda + 1)}.
\end{split}
\]
Notice that for integer values of $\m$ we see from the explicit solutions below that a logarithmic mode appears. In what follows we will consider only the case
$0 < |\m| < 2$ so $|\lambda| < \hf$, with $|\m| = 1$ the special point discussed above, so such logarithmic modes are not required. It would be straightforward to generalize
the linearized analysis to other values of $\lambda$, whilst for $\lambda < 0$ the corresponding dual operator is relevant and thus there is no obstruction to
carrying out a full non-linear analysis of the system.

Substituting the expansions into the on-shell action,
the second term in the expansion of the on-shell action $I_2$ was defined for $\m = 1$ in \eqref{eq:variationactionpoincare} and now becomes:
\be
I_{2,\lambda} = \frac{1}{32 \pi G_N}\int d^2 x \Big(h'_{ij} - \eta_{ij}\tr(h') - 2 \r \frac{1}{2\lambda + 1}\e_i^{\phantom{i}k}h''_{kj} - \frac{1}{(2\lambda + 1)}\e_i^{\phantom{i}k} h'_{kj}\Big) h^{ij}.
\ee
Substituting \eqref{eq:fgexpansiongeneralmu}, we find that this action is again divergent if $h_{(-2\lambda)}$ is nonzero and if $\lambda > 0$, with a leading piece of the form:
\be
I_{2,\lambda} = \frac{1}{32 \pi G_N \mu}\int d^2 x \Big( \frac{1}{2}\tr(h_{(0)}) \tilde R[h_{(-2\lambda)}] - 2\lambda h_{(2)ij}h^{ij}_{(-2\lambda)}  - \frac{1}{2} h_{(0)i}^k \del^i \del^j h_{(-2\lambda)jk} \Big)\r^{-\lambda} + \ldots
\ee
This term is cancelled precisely by adding $I_{2,\text{ct}}/(2\lambda + 1)$, where $I_{2,\text{ct}}$ is the counterterm action for $\m = 1$ defined in \eqref{eq:i2ct}. For $\lambda < 0$ there is no divergence but the counterterm action is then finite as well and we will continue to include it.

The variation of the total action $I_{2,\lambda,\text{tot}}= I_{2,\lambda} + I_{2,\text{ct}}/(2\lambda +1)$ is similar to \eqref{eq:variationpoinc}:
\begin{align}
\frac{\delta I_{2,\lambda,\text{tot}}}{\delta h^{ij}} &= \frac{1}{16 \pi G_N}\Big(h'_{ij} - \eta_{ij}\tr(h') + \frac{1}{2\lambda + 1}\Big[ - 2 \r \e_i^{\phantom{i}k}h''_{kj} - \e_i^{\phantom{i}k} h'_{kj} + \hf \tilde A_{ij}[h] \nn\\
&\qquad \qquad  + \frac{1}{4} \del^k \del_k h_{ij} - \frac{1}{4}\del_i \del^k h_{kj}\Big] \Big), \\
\frac{\delta I_{2,\lambda,\text{tot}}}{\delta h'^{ij}} &= \frac{1}{16 \pi G_N (2\lambda +1)}\r (\delta_i^k + \e_i^{\phantom{i}k})h'_{kj}.\nn
\end{align}
To obtain the one-point functions we follow the same reasoning as in section \ref{sec:oneptfunctions}. We have two independent variables, $h_{(0)ij}$ and $h_{(-2\lambda)ij}$, for which we define the corresponding CFT operators $T_{ij}$ and $X_{ij}$, with $T_{ij}$ again the energy-momentum tensor of the theory. To find their one-point functions, we first observe that:
\be
h_{(0)}^{ij} = \lim_{\r \to 0} (h^{ij} + \frac{1}{\lambda}h'^{ij}\r) \qquad \qquad  h_{(-2\lambda)}^{ij} = \lim_{\r \to 0} \left ( - \frac{1}{\lambda} h'^{ij} \r^{\lambda + 1} \right )
\ee
where we note that indices are raised with $\eta^{ij}$. From these expressions we find:
\be
\label{eq:vevsdefns}
\begin{split}
\vev{X_{ij}} &\equiv \frac{- 4\pi}{\sqrt{-g_{(0)}}}\frac{\delta I_{2,\lambda,\text{tot}}}{\delta h^{ij}_{(-2\lambda)}} = \lim_{\r \to 0} \Big(\lambda \r^{-1-\lambda} \frac{4\pi}{\sqrt{-g}}\frac{\delta I_{2,\lambda,\text{tot}}}{\delta h'^{ij}} - \r^{-\lambda}\frac{4\pi}{\sqrt{-g}}\frac{\delta I_{2,\lambda,\text{tot}}}{\delta h^{ij}} \Big)_L\\
\vev{T_{ij}} &\equiv \frac{4\pi}{\sqrt{-g_{(0)}}}\frac{\delta I_{2,\lambda,\text{tot}}}{\delta h^{ij}_{(0)}}  = \lim_{\r \to 0} \frac{4\pi}{\sqrt{-g}} \frac{\delta I_{2,\lambda,\text{tot}}}{\delta h^{ij}},
\end{split}
\ee
where the signs originate from the reasoning in appendix \ref{app:analyticcont}, plus an extra sign arising from the fact that $g^{ij} = \eta^{ij} - h^{ij}$. We inserted a factor of $4\pi$ in the definition of $X_{ij}$ for later convenience. After substitution of \eqref{eq:fgexpansiongeneralmu} these expressions lead to the following finite one-point functions:
\begin{align}
\vev{T_{ij}} &= \frac{1}{4 G_N}\Big\{(\d_i^k - \frac{1}{2\lambda + 1} \e_i^{\phantom{i}k}) h_{(2)kj} - \eta_{ij}\tr(h_{(2)}) + \frac{1}{2(2\lambda + 1)}\bar P_i^k \Big( \del^l \del_l h_{(0)kj} - \del_j \del^l h_{(0)kl}\Big)\Big\}, \nn \\
\label{eq:vevsgeneralmu} 
\vev{X_{ij}} &= \frac{\lambda(1+\lambda)}{2 G_N (2\lambda + 1)}(h_{(2+2\lambda)ij})_L .
\end{align}
Symmetrization in $i$ and $j$ is again understood in these expressions.


\subsection{Two-point functions}
Just as in section \ref{sec:exactsolutions}, we can use the equations \eqref{eq:Tbessel} and \eqref{eq:UpVp} (with the $K_\m$ choice for the Bessel function) to find exact solutions to the linearized equations of motion. Asymptotically, they behave as follows:
\be
\begin{split}
h_{z\bar z} &= h_{(0) z\bar z} + \r h_{(2) z\bar z} + \frac{1}{2(\lambda-1)(2\lambda +1)}\del^2 h_{(-2\lambda )\bar z \bar z} \r^{1-\lambda} + \ldots,\\
h_{\bar z\bar z} &= h_{(-2\lambda)\bar z \bar z}\r^{-\lambda} + h_{(0) \bar z\bar z} + \frac{1}{2(\lambda -1)} \bar \del \del h_{(-2\lambda)\bar z \bar z} \r^{1-\lambda} + \frac{\bar \del}{\del}h_{(2) z\bar z} \r + \ldots,\\
h_{zz} &= h_{(0) zz} + \frac{\del}{\bar \del} h_{(2) z\bar z} \r + \frac{2^{-4\lambda +2}\lambda}{(\lambda + 1)}\frac{\G(-2\lambda -1)}{\G(2\lambda + 1)}  q^{4\lambda -2}\del^4 h_{(-2\lambda)\bar z \bar z}\r^{\lambda + 1} + \ldots,
\end{split}
\ee
with same trace condition as was given for $\m = 1$ in \eqref{eq:linearizedtrace},
\be
h_{(2) z\bar z} = - \frac{1}{2} \del^2 h_{(0) \bar z\bar z} -\frac{1}{2} \bar \del^2 h_{(0) zz} + \bar \del \del h_{(0) z\bar z},
\ee
and integration constants $h_{(0) \bar z\bar z},h_{(0) zz},h_{(0) z\bar z}$ and $h_{(-2\lambda)\bar z \bar z}$; these are as anticipated the sources for the dual operators.

We can substitute this solution in \eqref{eq:vevsgeneralmu} to find the one-point functions:
\be
\begin{split}
\vev{X_{zz}} &= \frac{2^{-4\lambda + 1} \lambda^2}{G_N} \frac{\G(-2\lambda -1)}{\G(2\lambda + 2)}  q^{4\lambda -2}\del^4 h_{(-2\lambda)\bar z \bar z}\\
\vev{T_{\bar z \bar z}} &= \frac{2 \lambda + 2}{4 G_N (2 \lambda + 1)} \frac{\bar \del}{\del}h_{(2) z\bar z} + \text{local}\\
\vev{T_{z \bar z}} &= \text{local}\\
\vev{T_{zz}} &= \frac{2\lambda}{4 G_N (2\lambda + 1)}\frac{\del}{\bar \del}h_{(2) z\bar z} .
\end{split}
\ee
From these expressions we obtain the following nonvanishing two-point functions:
\be
\label{eq:twoptgeneralmu}
\begin{split}
\vev{T_{\bar z \bar z}(z,\bar z)T_{\bar z \bar z}(0)} &= \frac{i \pi}{2 G_N} \frac{\lambda + 1}{2 \lambda + 1} \frac{\bar \del^3}{\del} \d^2(z,\bar z) =  \frac{3}{ 2 G_N}\frac{\lambda + 1}{2\lambda + 1} \frac{1}{\bar z^4},\\
\vev{T_{zz}(z,\bar z)T_{zz}(0)} &= \frac{i \pi}{2 G_N} \frac{\lambda}{2 \lambda + 1} \frac{\del^3}{\bar \del} \d^2(z,\bar z) =  \frac{3}{ 2 G_N}\frac{\lambda}{2\lambda + 1} \frac{1}{z^4},\\
\vev{X_{zz}(z,\bar z) X_{zz}(0)} &= i \frac{4 \pi}{\sqrt{-g_{(0)}}} \frac{\d}{\d h_{(-2\lambda)}^{zz}(z,\bar z)} \vev{X(0)} = 2 \pi i \frac{\d}{\d h_{(-2\lambda)\bar z \bar z}(z,\bar z)} \vev{X_{zz}(0)}\\
& = \frac{i \pi 2^{-4\lambda + 2} \lambda^2}{G_N} \frac{\G(-2\lambda -1)}{\G(2\lambda +2)} q^{4\lambda -2}\del^4 \d^2(z,\bar z)\\
& = \frac{-1}{2 G_N} \frac{\lambda (\lambda + 1)(2\lambda +3)}{2\lambda +1}\frac{1}{z^{2\lambda + 4}\bar z^{2\lambda} },
\end{split}
\ee
where the computation of the two-point function of the energy-momentum tensor is completely analogous to the previous section and we used the identity
\eqref{eq:fourierqa}. Comparing now with \eqref{eq:clcr} we read off that:
\be \label{eq:centralcharges}
(c_L, c_R) = \frac{3}{G_N}  \Big(\frac{\lambda}{2\lambda + 1},\frac{\lambda + 1}{2\lambda + 1}\Big) =  \frac{3}{2G_N}  \Big( 1 - \frac{1}{\m},1+ \frac{1}{\m}\Big)
\ee
and from the last line in \eqref{eq:twoptgeneralmu} we also find that $X$ has weights $(h_L,h_R) = (2+ \lambda,\lambda) = \hf (\m + 3,\m - 1)$. Both expressions agree with \cite{Li:2008dq}.

\paragraph{The limit $\lambda \to 0$ and logarithmic CFT\\}

As $\lambda \to 0$, we find that the $\vev{TT}$-correlators return to the values given in section \ref{sec:twoptfunctions}. On the other hand, the $\vev{XX}$-correlator vanishes, but we also find that the definitions for $X_{zz}$ and $T_{zz}$ as given in \eqref{eq:vevsdefns} coincide in this limit (up to a sign). To remedy this we can introduce a new field,
\be
t_{zz} = - \frac{1}{\lambda} X_{zz} - \frac{1}{\lambda}T_{zz},
\ee
after which we can take $\lambda \to 0$ in \eqref{eq:vevsdefns} and obtain \eqref{eq:vevsdefnsm1} (up to a sign from the fact that $g^{ij} = \eta^{ij} - h^{ij}$). We obtain for the nonzero two-point functions:
\be
\label{eq:tcorr}
\begin{split}
\vev{t_{zz}(z,\bar z) T_{zz}(0)} &= - \frac{3}{2G_N}\frac{1}{2\lambda + 1} \frac{1}{z^4} = \frac{- 3/(2G_N)}{z^4} + \ldots\\
\vev{t_{zz}(z,\bar z)t_{zz}(0)} &= \frac{B_m + 3/(G_N) \log(m^2 |z|^2)}{z^4} + \ldots
\end{split}
\ee
where the dots represent terms that vanish as $\lambda \to 0$. These are exactly the same correlators as in section \ref{sec:twoptfunctions}.
The term $B_m$ can again be removed by a redefinition of $t_{zz}$ and from \eqref{eq:tcorr} we again see that $b = - 3/G_N$.

In appendix \ref{sec:applcft} we discuss the degeneration of a CFT to a logarithmic CFT as $c_L \rightarrow 0$ following
Kogan and Nichols \cite{Kogan:2002mg}. Their
$c_L \rightarrow 0$ limit is precisely the same limit as taken here, i.e. the logarithmic partner of the stress energy tensor originates from another primary operator
whose dimension approaches $(2,0)$ in the $c_{L} \rightarrow 0$ limit. Given such a limiting procedure, the anomaly $b$ is
obtained by inverting the relation
between $\lambda$ (which is the right-moving weight of $X$) and $c_L$ given above and using \eqref{eq:b} in appendix \ref{sec:applcft}. This results
in $b = - \lim_{c_R \to 0} c_L/\lambda(c_L) = - 3/G_N$ and thus agrees with \eqref{eq:tcorr}. Note that there are other distinct approaches to
taking a $c \rightarrow 0$
limit, see \cite{Flohr:2005dr} for a review, but it is the Kogan-Nichols approach which is realized holographically here.

\paragraph{Energy computations\\}

In Lorentzian signature and in global coordinates, the insertions of the operators $X_{zz}$, $T_{zz}$ or $T_{\bar z \bar z}$ in the infinite past creates the massive, left-moving or right-moving graviton states discussed in \cite{Li:2008dq}. In \cite{Li:2008dq} the energy of these states was computed in the bulk and we are now able to give a CFT interpretation of their results.

For the states created by the operators $X_{zz},T_{zz},T_{\bar z \bar z}$, the equations (70)-(72) in \cite{Li:2008dq} give energies of the form:
\be
\label{eq:energies}
\begin{split}
X_{zz}: \qquad \qquad E_M &= \frac{-1}{8 G_N}(\m - \frac{1}{\m}) (h_L + h_R) \Big[ \ldots \Big], \\
T_{zz}: \qquad \qquad E_L &= \frac{-1}{4 G_N}(-1 + \frac{1}{\m}) \Big[\ldots\Big], \\
T_{\bar z \bar z}: \qquad \qquad E_R &= \frac{-1}{4 G_N} (-1 - \frac{1}{\m})  \Big[\ldots\Big]. \\
\end{split}
\ee
The expressions in square brackets are positive, but their exact value depends on the normalization of the solutions to the linearized equations of motion in \cite{Li:2008dq} and is therefore arbitrary. We can thus only compare the overall sign of the energies \eqref{eq:energies} with our results. Notice that we put in an extra factor of the left- plus right-moving weight from each operator, which for $T_{zz}$ and $T_{\bar z \bar z}$ are just factors of 2; in \cite{Li:2008dq} such factors comes from a time derivative of the bulk modes and we will see similar factors appearing below.

Following the usual CFT logic, we may obtain the energies of a state by computing three-point functions. For example, for the massive mode we need to compute
\be
\vev{X_{zz}| T_{zz}(z) |X_{zz}},
\ee
with
\be
|X_{zz}\> = X_{zz}(0,0) |0\>, \qquad \qquad \<X_{zz}| = \lim_{z,\bar z \to \infty} \< 0| X_{zz}(z,\bar z) z^{2\lambda + 4} \bar z^{2\lambda}.
\ee
The usual Ward identity:
\be
\vev{X_{zz}(z_1) T_{zz}(z) X_{zz}(z_2)} = \sum_{i \in \{1,2\}} \Big( \frac{h_L}{(z - z_i)^2} + \frac{1}{z - z_i} \frac{\del}{\del z_i} \Big) \vev{X_{zz}(z_1) X_{zz}(z_2)}
\ee
results in:
\be
\vev{X_{zz}| T_{zz}(z) |X_{zz}} =  \frac{C_X h_L}{z^2},
\ee
where $C_X$ is the normalization of the $\vev{XX}$-correlator,
\be
\vev{X_{zz}(z,\bar z)X_{zz}(0)} = \frac{C_X}{z^{4 + 2\lambda} \bar z^{2\lambda}},
\ee
\[
C_X = \frac{-1}{2 G_N} \frac{\lambda (\lambda + 1)(2\lambda +3)}{2\lambda +1} = \frac{- 1}{8 G_N} (\m - \frac{1}{\m}) (\m + 2).
\]
Note that the magnitude (but not the sign) of $C_X$ can change by changing the normalization of the 
operator $X$. This is the counterpart of the arbitrariness of the quantities in the 
square brackets of \eqref{eq:energies} due to the normalization
ambiguity of the solutions to the linearized equations.

By using the Virasoro algebra one may also obtain that:
\be
\vev{T_{zz}| T_{zz}(z) |T_{zz}} = \vev{0|L_{2} \sum_{m \in \mathbb Z} L_m z^{-m -2}L_{-2}|0} = \frac{c_L}{z^2},
\ee
with $c_L$ the left-moving central charge defined in \eqref{eq:centralcharges}. The computation involving $T_{\bar z \bar z}$ is completely analogous, and of course the mixed three-point functions involving $T_{zz}$ and $T_{\bar z \bar z}$ vanish. To transfer these results to the cylinder we use the conformal transformation:
\be
z = \exp(i w),
\ee
whose Schwarzian derivative is $1/2$. We then find the following three-point functions on the cylinder:
\be
\label{eq:unnormenergies}
\begin{split}
\vev{X_{ww}| T_{ww}(w)+ T_{\bar w \bar w}(\bar w) - \frac{c_L + c_R}{24} |X_{ww}} &=
C_X (h_L + h_R) \\&=  \frac{- 1}{8 G_N} (\m - \frac{1}{\m}) (h_L + h_R) (\m + 2), \\
\vev{T_{ww}| T_{ww}(w)+ T_{\bar w \bar w}(\bar w)  - \frac{c_L + c_R}{24} |T_{ww}} &= c_L =  \frac{3}{2 G_N}(1 - \frac{1}{\m}),\\
\vev{T_{\bar w \bar w}| T_{ww}(w)+ T_{\bar w \bar w}(\bar w) - \frac{c_L + c_R}{24} |T_{\bar w \bar w}} &= c_R = \frac{3}{2 G_N}(1 + \frac{1}{\m}).
\end{split}
\ee

Let us now compare these results with \cite{Li:2008dq}. Notice first of all that the zero-point of energy in that paper is that of global AdS, which is why we explicitly subtracted the Casimir energy in the above expressions. Comparing now \eqref{eq:unnormenergies} with \eqref{eq:energies} we indeed find the same structure and precisely the same signs. The computations are therefore in agreement.

Finally, notice that in a CFT one usually divides the expressions in \eqref{eq:unnormenergies} by the norm of the state (\emph{e.g.} $\vev{X_{zz}|X_{zz}}$) to obtain energies that are precisely equal to the conformal weights of the operators creating the state. On the other hand, the energies computed using bulk methods as in \cite{Li:2008dq} are the unnormalized energies of \eqref{eq:unnormenergies} and therefore extra signs may arise if a state has a negative norm. This explains the sign difference between the conformal weights and the energies found in \cite{Li:2008dq}.

\section{Conclusions}

By implementing the AdS/CFT dictionary for topologically massive
gravity, we were able to provide further evidence for its duality at
$\m = 1$ to a logarithmic conformal field theory. The expressions for
the two-point functions indicate problems with unitarity and positivity
as we find zero-norm states at $\m =1$, negative-norm states at $\m \neq
1$ and negative conformal weights at $\m < 1$. It therefore seems
problematic to consider the full TMG as a fundamental theory, but this
duality could nonetheless have interesting applications to
condensed matter systems. For example, $c=0$ LCFTs arise in the description
of critical systems with quenched disorder and in several other contexts.

One may try to restrict to the right-moving sector of the
theory \cite{Maloney:2009ck}, which could yield a consistent chiral
theory. In order for this sector to decouple a necessary requirement is that the
$\vev{t  \bar T \bar T}$ three-point function should vanish. This
was shown to be the case in the discussion of \cite{Kogan:2002mg},
see their equation (42), and their analysis can be adapted to 
the case of interest, namely when only $c_L \to 0$,
leading to the same result. This 
suggests that one can indeed truncate to the right-moving sector, but 
it would
be interesting to extend our analysis and verify the vanishing of 
this 3-point function by a bulk computation. 

One may also perform a holographic analysis for the `warped'
solutions found in \cite{Anninos:2008fx}. The asymptotics
in these cases are discussed in appendix \ref{app:warped} and
indicate qualitatively different UV behavior for the dual field theory; it would 
be interesting to extend the holographic setup to this class of solutions.
A similar procedure could also be followed to analyze the `new
massive gravity' of \cite{Bergshoeff:2009hq} around AdS
solutions. This would allow us to find out more about the possible dual
CFTs.

\section*{Acknowledgments}

This work is part of the research program of the 'Stichting voor
Fundamenteel Onderzoek der Materie (FOM)', which is financially
supported by the `Nederlandse Organisatie voor Wetenschappelijk
Onderzoek (NWO)'. The authors acknowledge support from NWO, KS via a
Vici grant, MMT via the Vidi grant ``Holography, duality and time
dependence in string theory'' and BvR via an NWO Spinoza grant. KS and
MMT would like to thank GGI in Florence and the Aspen Center of
Physics for hospitality during the final stages of this work.

\appendix
\section{Derivation of the equations of motion}
\label{sec:appeom}
In this appendix we derive the equations of motion in Fefferman-Graham coordinates, where the metric has the form
\be
\label{eq:fgmetricapp}
ds^2 = \frac{d\r^2}{4\r^2} + \frac{1}{\r} g_{ij}(x,\r)dx^i dx^j.
\ee
In this section we raise indices using $g^{ij}$ and the covariant derivative $\cdel_i$ and the two-dimensional antisymmetric tensor $\e_{ij}$ are also defined using $g_{ij}$. In the metric \eqref{eq:fgmetricapp} the nonzero connection coefficients are:
\begin{align}
\label{eq:connections}
\G_{\r\r}^\r &= - \frac{1}{\r} & \G_{\r j}^i &= - \frac{1}{2\r} g^i_j + \hf (g^{-1} g')^i_j\\
\G_{ij}^\r &= 2 g_{ij} - 2 \rho g'_{ij} & \G_{jk}^i &= \G_{jk}^i(g)\,,
\end{align}
where the index $\r$ now denotes the coordinate $\r$ and a prime denotes radial derivative. The curvature tensor becomes:
\bea
R_{\r ij}{}^k(G) &=& \frac{1}{2} g^{kl} \left(\nabla_l g'_{ij}
- \nabla_j g'_{il} \right), \nn \\
\label{eq:riemanns}
R_{i \r j}{}^\r(G) &=& - 2 \r \left(g''_{ij} - \frac{1}{2} (g' g^{-1} g')_{ij} \right)
- \frac{1}{\r} g_{ij}, \\
R_{ijk}{}^l(G) &=& R_{ijk}{}^l(g) + \left(
\frac{1}{\r} g_i^l g_{jk} +g_j^l g'_{ik} + g^{ml} g_{ik}  g'_{mj}
+ \r g^{lm} g'_{im}  g'_{jk} - (i \leftrightarrow j) \right), \nn
\eea
The Einstein part of the equation of motion, $R_{\m \n} + 2 G_{\m \n}$, is given by:
\be
\begin{split}
\label{eq:einstein}
R_{\r\r}(G)  + 2 G_{\r\r} &= - \hf \tr(g^{-1}g'') + \qt \tr(g^{-1}g' g^{-1}g'),\\
R_{i \r}(G) + 2 G_{i \r} &= \hf \cdel^j g'_{ji} - \hf \cdel_i \tr(g^{-1}g'),\\
R_{ij}(G) + 2 G_{ij} &= \hf R(g) g_{ij} + g_{ij} \tr(g^{-1}g') + \r \big[-2 g''_{ij} - g'_{ij} \tr(g^{-1}g') + 2 (g' g^{-1}g')_{ij} \big],
\end{split}
\ee
where we used that in two dimensions
\be
R_{ijkl} = \frac{1}{2} R [g_{ik}g_{lj} - (l \leftrightarrow k)], \qquad \qquad R_{ik} = \frac{1}{2}R g_{ik}\,.
\ee
The trace equation $R = -6$ now becomes:
\be
\label{eq:appR}
- 4 \r \tr(g^{-1}g'') + 3 \r \tr(g^{-1}g' g^{-1}g') - \r [\tr(g^{-1}g')]^2 + R(g) + 2 \tr(g^{-1}g') = 0.
\ee
We use $\e^{\rho ij} = 2 \rho^2 \e^{ij}$ to relate the three- and two-dimensional $\epsilon$-tensors. For the Cotton tensor $C_{\m \n}$ defined in \eqref{cotton} we then find:
\be
\label{eq:cotton}
\begin{split}
C_{\r\r} &= \qt \e^{ij}\Big( \cdel_i \cdel^k g'_{kj} + 2 \r(g'' g^{-1}g')_{ji} \Big),\\
C_{\r i} &= \hf \e^{jk}\Big( \hf g_{ik} \cdel_j R - 2 \r \cdel_j g''_{ik} - \r \tr(g^{-1}g') \cdel_j g'_{ik} + 2 \r \cdel_j (g'g^{-1}g')_{ik} \\&\qquad \qquad - (g_{ij} - \r g'_{ij}) \cdel^l g'_{lk}\Big),\\
C_{i\r} &= \e_i^{\phantom{i} k} \Big( - \r \cdel^l g''_{lk} - \frac{1}{4}\r \cdel_k \tr( g^{-1}g'g^{-1}g') + \hf \r (g^{-1}g')^j_k \cdel^l g'_{lj} + \r (g^{-1}g')^j_l\cdel^l g'_{jk} \\ &\qquad\qquad + \frac{1}{2} \cdel_k \tr(g^{-1}g') - \frac{1}{2} \cdel^j g'_{jk}\Big),\\
C_{ij} &= 2 \r \e_i^{\phantom{i} k}\Big( g_{jk}[ - \hf R' - \frac{1}{4\r} R  - \frac{1}{2\r}\tr(g^{-1}g') + \hf \tr(g^{-1}g'g^{-1}g')] - \qt R g'_{jk} \\
&\qquad \qquad + \hf \cdel_k \cdel^m g'_{mj} - \hf \cdel_k \cdel_j[\tr(g^{-1}g')] + 2 \r g'''_{jk} + g''_{kj}[3 + \r \tr(g^{-1}g')] \\&\qquad \qquad  + g'_{kj}[\tr(g^{-1}g') + \r (\tr(g^{-1}g'))' -\r \tr(g^{-1}g'') + \hf \r \tr(g^{-1}g'g^{-1}g')] \\ &\qquad \qquad + (g'g^{-1}g')_{kj} [-3 - \hf \r \tr(g^{-1}g')] - 3 \r (g'' g^{-1}g')_{kj} - 2\r (g' g^{-1} g'')_{kj} \\&\qquad \qquad + 3 \r (g' g^{-1}g' g^{-1}g')_{kj}
\Big). \\
\end{split}
\ee
With these expressions we indeed find that $C^\mu_\mu = 0$, $C_{\r i} = C_{i\r}$ and $C_{ij} = C_{ji}$. To verify this we used the Cayley-Hamilton identity,
\be
\label{eq:cayleyhamilton}
\hf g_{jl} \Big( [\tr(g^{-1}g')]^2 - \tr(g^{-1}g' g^{-1}g')\Big) +  (g' g^{-1} g')_{jl} - g'_{jl} \tr(g^{-1}g') = 0\,,
\ee
the radial derivative of the two-dimensional Ricci tensor,
\be
R'_{ik} = \hf \Big( \cdel^l \cdel_i g'_{kl} + \cdel^l \cdel_k g'_{il} - \cdel^a \cdel_a g'_{ik} - \cdel_i \cdel_k \tr(g^{-1}g')\Big)\,,
\ee
as well as the identity for the two-dimensional $\e$-symbol,
\be
\e_{ij}\e_{kl} = - g_{ik}g_{jl} + g_{il} g_{jk}\,.
\ee
As $C_{ij}$ is symmetric, we can also rewrite it as $\hf(C_{ij} + C_{ji})$ which allows us to drop the term proportional to $\e_i^{\phantom{i}k} g_{kj}$. This, the expression for $R$ given in \eqref{eq:appR}, and further application of the Cayley-Hamilton theorem eventually give:
\be
\begin{split}
C_{ij} &= \r \e_i^{\phantom{i} k}\Big( \hf \cdel_k \cdel^m g'_{mj} - \hf \cdel_k \cdel_j[\tr(g^{-1}g')] + 2 \r g'''_{jk} + g''_{kj}\Big[3 + \r \tr(g^{-1}g')\Big] \\&\qquad  + g'_{kj}\Big[-\frac{3}{2}\tr(g^{-1}g') + \frac{3}{4} \r [\tr(g^{-1}g')]^2 -\r \tr(g^{-1}g'') + \frac{7}{4} \r \tr(g^{-1}g'g^{-1}g')\Big] \\ &\qquad - 3 \r (g'' g^{-1}g')_{kj} - 2\r (g' g^{-1} g'')_{kj} \Big)  + i \leftrightarrow j\,.
\end{split}
\ee
Combining the above expressions \eqref{eq:einstein} and \eqref{eq:cotton} leads to the full equations of motion which are given by:
\begin{align}
&- \hf \tr(g^{-1}g'') + \qt \tr(g^{-1}g' g^{-1}g') + \frac{1}{4\m}\e^{ij}\Big( \cdel_i \cdel^k g'_{kj} + 2 \r(g'' g^{-1}g')_{ji} \Big) = 0 ,\nn \\
&\hf \cdel^j g'_{ji} - \hf \cdel_i \tr(g^{-1}g') + \frac{1}{2\m}\e^{jk}\Big( \hf g_{ik} \cdel_j R + g_{ik}\cdel^l g'_{lj} \nn \\
&\qquad + \r \Big[ - 2 \cdel_j g''_{ik} - \tr(g^{-1}g') \cdel_j g'_{ik} + 2 \cdel_j (g' g^{-1}g')_{ik}  + g'_{ij} \cdel^l g'_{lk}\Big] \Big) = 0,\\
& \Big(\tr(g^{-1}g'') - \frac{3}{4} \tr(g^{-1}g' g^{-1}g') + \qt [\tr(g^{-1}g')]^2 \Big)g_{ij}  - g''_{ij} - \hf g'_{ij} \tr(g^{-1}g') + (g' g^{-1}g')_{ij}  \nn \\
&\qquad + \frac{1}{\m}\e_i^{\phantom{i} k}\Big( \hf \cdel_k \cdel^m g'_{mj} - \hf \cdel_k \cdel_j[\tr(g^{-1}g')] + 2 \r g'''_{jk} + g''_{kj}\Big[3 + \r \tr(g^{-1}g')\Big] \nn \\
& \qquad  + g'_{kj}\Big[-\frac{3}{2}\tr(g^{-1}g') + \frac{3}{4} \r [\tr(g^{-1}g')]^2 -\r \tr(g^{-1}g'') + \frac{7}{4} \r \tr(g^{-1}g'g^{-1}g')\Big] \nn \\ &\qquad - 3 \r (g'' g^{-1}g')_{kj} - 2\r (g' g^{-1} g'')_{kj} \Big)  + i \leftrightarrow j = 0,\nn
\end{align}
where we emphasize that the symmetrization in the last equation concerns all the terms. We can use the $(\r\r)$ equation of motion to simplify the $(ij)$ equation of motion to:
\begin{align}
& \Big(\hf \tr(g^{-1}g'') - \hf \tr(g^{-1}g' g^{-1}g') + \qt [\tr(g^{-1}g')]^2 \Big)g_{ij}  - g''_{ij} - \hf g'_{ij} \tr(g^{-1}g') + (g' g^{-1}g')_{ij}  \nn \\& \qquad  + \frac{1}{\m}\e_i^{\phantom{i} k}\Big( \qt \cdel_k \cdel^m g'_{mj} + \qt \cdel_j \cdel^m g'_{mk} - \hf \cdel_k \cdel_j[\tr(g^{-1}g')] + 2 \r g'''_{jk} + g''_{kj}\Big[3 + \r \tr(g^{-1}g')\Big] \nn \\&\qquad + g'_{kj}\Big[-\frac{3}{2}\tr(g^{-1}g') + \frac{3}{4} \r [\tr(g^{-1}g')]^2 -\r \tr(g^{-1}g'') + \frac{7}{4} \r \tr(g^{-1}g'g^{-1}g')\Big] \nn \\ &\qquad - \frac{5}{2} \r (g'' g^{-1}g')_{kj} - \frac{5}{2}  \r (g' g^{-1} g'')_{kj} \Big)  + i \leftrightarrow j = 0.
\end{align}
If we use the first radial derivative of \eqref{eq:cayleyhamilton} we can simplify this further to:
\begin{align}
\label{eq:ijeom}
& \Big(\frac{1}{2}\tr(g^{-1}g'') - \qt [\tr(g^{-1}g')]^2 \Big)g_{ij}  - g''_{ij} + \hf g'_{ij} \tr(g^{-1}g')  \\&\qquad + \frac{1}{\m}\e_i^{\phantom{i} k}\Big( \qt \cdel_k \cdel^m g'_{mj} + \qt \cdel_j \cdel^m g'_{mk} - \hf \cdel_k \cdel_j[\tr(g^{-1}g')] + 2 \r g'''_{jk} + g''_{kj}[3 -\frac{3}{2} \r \tr(g^{-1}g')] \nn \\&\qquad  + g'_{kj}[- \frac{3}{2}\tr(g^{-1}g') + \frac{3}{4} \r [\tr(g^{-1}g')]^2 - \frac{7}{2} \r \tr(g^{-1}g'') + \frac{7}{4}\r \tr(g^{-1}g'g^{-1}g')]
\Big) + i \leftrightarrow j = 0.\nn
\end{align}
We can use the equation of motion to rewrite the Riemann tensor as:
\be
R_{\a \b\g \d}[G]  = G_{\a\d}G_{\b \g} - G_{\a\g}G_{\b\d}
- \Big( \big[ \frac{1}{\mu}G_{\a\g} C_{\b \d} - (\a \leftrightarrow \b) \big]
- (\g \leftrightarrow \d) \Big),
\ee
Using then \eqref{eq:riemanns} for the Riemann tensor in Fefferman-Graham coordinates we obtain:
\be
\begin{split}
&-2 g''_{ij} + (g'g^{-1}g')_{ij} + \frac{4}{\m}g_{ij}C_{\r\r} + \frac{1}{\m \r}C_{ij} = 0,\\
&\hf \Big( \cdel_k g'_{ij} - \cdel_j g'_{ik} \Big) = \frac{1}{\m}(g_{ij}C_{\r k} - g_{ik}C_{\r j}),\\
&\hf\Big(g_{ik}g_{jl} - g_{il}g_{jk} \Big)\Big( -2 \tr(g^{-1}g') + \r [\tr(g^{-1}g')]^2 - \r \tr(g^{-1}g'g^{-1}g')\Big) \\ &\qquad + \Big( g_{jl}g'_{ik} + g_{ik}g'_{jl} + \r g'_{il} g'_{jk} - (i \leftrightarrow j) \Big) =0.
\end{split}
\ee
Taking the trace $g^{ik} R_{ijkl}$ of the last equation results again in the Cayley-Hamilton identity \eqref{eq:cayleyhamilton}. This is also the equation that one obtains from the first equation by eliminating $C_{ij}$ and $C_{\r\r}$ using the equations of motion. On the other hand, the second of these equations can alternatively be written as:
\be
\label{eq:secondrijkl}
\begin{split}
& (g^{kj} - \m \e^{kj}) \cdel_k g'_{ij} - \cdel_i \Big( \tr(g^{-1}g') + \hf \r \tr(g^{-1}g'g^{-1}g') - \r [\tr(g^{-1}g')]^2\Big) \\
& \qquad \qquad + 2 \r \cdel^n \Big(g''_{in} - \tr(g^{-1}g') g'_{in}\Big) + \r (g^{-1}g')^k_i \cdel^l g'_{kl} = 0\,.
\end{split}
\ee

\section{Wick rotation}
\label{app:analyticcont}
Given a Lorentzian theory, the most straightforward way to find the corresponding action in Euclidean signature is to use a complex diffeomorphism:
\be
t = -i\tau.
\ee
After this diffeomorphism (or a similar one using a different coordinate system) the metric generally becomes positive definite and one has to be careful about the definition of the square root in the metric determinant. The signs work out correctly if we define $\sqrt{-1} = -i$ \cite{Skenderis:2008dg}. As in any coordinate system, the antisymmetric tensor is still defined such that $\sqrt{-G} \e^{0 1 2} = 1$ with $x^0$ now the $\tau$-direction. Because of the volume element the $\e$-tensor is now complex and to comply with standard notation we make this explicit by writing $- i \epsilon^{\lambda \m \n} = \hat \epsilon^{\lambda \m \n}$, where $\hat \epsilon^{\lambda \m \n}$ is the standard antisymmetric tensor in Euclidean coordinates which is defined such that $\sqrt{G} \hat \e^{0 1 2} = 1$.

As for the action of the theory, we find that the diffeomorphism results in $iS_L \to -S_E$ with $S_E$ the standard Euclidean action. In our case, \eqref{eq:bulkaction} becomes:
\be
\begin{split}
i S_L &= - \frac{1}{16 \pi G_N} \int d^3 x \, \sqrt{G}(- R + 2 \Lambda) \\ &\qquad \qquad + \frac{i}{32 \pi G_N \mu} \int d^3 x \, \sqrt{G} \hat \e^{\lambda\m\n} \Big( \G_{\lambda \s}^\r \del_\m \G_{\r \n}^\s + \frac{2}{3}\G_{\lambda \s}^\r \G_{\m \t}^{\s}\G_{\n \r}^\t \Big).
\end{split}
\ee
Notice that the implicit metric determinant present in the $\epsilon$-symbol cancels the one in the volume element and there is no sign change for the Chern-Simons term. From this action, we see that a convenient way to determine the Euclidean equations of motion is to replace everywhere
\be
\e^{\lambda \m \n} \to i \hat \epsilon^{\lambda \m \n}, \qquad \qquad \e^{ij} \to i \hat \e^{ij}.
\ee
With these replacements the equations of motion become complex, and so do the linearized solutions we find in the main text, but this is not a problem, see \cite{Skenderis:2008dg} for a more extended discussion of this point.

When using component equations, the conversion between Euclidean and Lorentzian signature is most easily done by introducing lightcone coordinates on the Lorentzian side:
\be
u = x + t, \qquad \qquad  v = x - t.
\ee
In these coordinates the metric becomes:
\be
ds^2 = du dv
\ee
and we fix the sign of the $\e$-tensor such that $\e_{uv} = -\hf$. The passage to Euclidean signature is then implemented by defining complex coordinates:
\be
z = x + i\tau, \qquad \qquad \bar z = x - i\tau,
\ee
after which the metric $ds^2 = d\tau^2 + dx^2$ becomes:
\be
ds^2 = dz d\bar z.
\ee
The metric determinant in complex coordinates becomes negative again and therefore $\hat \e^{ij}$ is complex and $\e^{ij}$ is real. We deduce that the component equations in Euclidean signature can be obtained by the simple replacement
\be
v \to z, \qquad \qquad u \to \bar z,
\ee
in the Lorentzian equations of motion, without any modification of the $\epsilon$-tensor.

Incidentally, notice that the operators:
\be
P_i^k = \hf (\d_i^k + \e_i^{\phantom{i}k}), \qquad \qquad \bar P_i^k = \hf (\d_i^k - \e_i^{\phantom{i}k}),
\ee
take the following form in lightcone coordinates:
\be
\begin{pmatrix}
P_u^u & P_u^v\\
P_v^u & P_v^v
\end{pmatrix}
=
\begin{pmatrix}
0 & 0 \\ 0 & 1\\
\end{pmatrix}
\qquad \qquad
\begin{pmatrix}
\bar P_u^u & \bar P_u^v\\
\bar P_v^u & \bar P_v^v
\end{pmatrix}
=
\begin{pmatrix}
1 & 0 \\ 0 & 0\\
\end{pmatrix}
\ee
so that, if for example $P_i^k b_{(0)kj} = 0$ and $b_{(0)i}^i = 0$ then only the $b_{(0)uu}$ component can be nonzero. From the above reasoning it follows that these operators take the same form in complex coordinates and therefore only $b_{(0)\bar z \bar z}$ can be nonzero.

\paragraph{Signs in correlation functions\\}

Our conventions are such that on a Euclidean background metric $g_{ij}$ the energy-momentum tensor is defined as:
\be
\label{eq:euclT}
T_{E,ij} = \frac{4\pi}{\sqrt{g}} \frac{\d S_E}{\d g^{ij}}.
\ee
Notice that we functionally differentiate with respect to the inverse metric. When we analytically continue back to Lorentzian signature, the definition on the right-hand side changes. Namely, from the above discussion it follows that $S_{E} = - i S_L$ and $\sqrt{g} = i \sqrt{-g}$, so in Lorentzian signature
\be
\label{eq:lorT}
T_{L,ij} = - \frac{4\pi}{\sqrt{-g}} \frac{\d S_L}{\d g^{ij}}.
\ee
In terms of the generating functional of connected correlation functions, $W = \log(Z)$, we find that:
\be
\begin{split}
T_{E,ij} &= - \frac{4\pi}{\sqrt{g}} \frac{\d W_E}{\d g^{ij}},  \qquad \qquad T_{L,ij} = i \frac{4\pi}{\sqrt{-g}} \frac{\d W_L}{\d g^{ij}}.
\end{split}
\ee
These expressions lead to the following identity that we use in the main text:
\be
\vev{ T_{ij} \ldots}_g = i \frac{4\pi}{\sqrt{- g}} \frac{\d}{\d g^{ij}} \vev{\ldots}_g
\ee
where $\vev{\ldots}_g$ is an arbitrary correlator in the background metric $g_{ij}$. Notice that this expression holds irrespective of the signature of the metric, provided we define the square root as above.

Now for general correlation functions of an operator $\op$, we customarily define the source-operator coupling in Euclidean signature as:
\be
- \int d^2 x \sqrt{-g} \, \phi_{E} \cdot \op_E,
\ee
with $\phi_{E}$ the Euclidean source and the dot denoting various possible index contractions. Using once more the above conventions, we find that in Lorentzian signature the coupling becomes:
\be
- i \int d^2 x \sqrt{-g}\, \phi_{L} \cdot \op_L,
\ee
and therefore
\be
\vev{\op_E} = - \frac{1}{\sqrt{g}} \frac{\d W_E}{\d \phi_{E}}, \qquad \qquad \vev{\op_L} = i \frac{1}{\sqrt{-g}} \frac{\d W_L}{\d \phi_{L}}.
\ee
This results in the general expression in terms of correlation functions:
\be
\vev{\op \ldots}_{\phi} = i \frac{1}{\sqrt{-g}} \frac{\d}{\d \phi} \vev{\ldots}_{\phi}.
\ee

In the context of AdS/CFT, $W_E \sim - S_E$ and $W_L \sim i S_L$ with $S_E$ and $S_L$ the Euclidean and the Lorentzian on-shell bulk action, respectively. This leads to:
\be
\vev{\op_E} = \frac{1}{\sqrt{g}} \frac{\d S_E}{\d \phi_{E}}, \qquad \qquad \vev{\op_L} = - \frac{1}{\sqrt{-g}} \frac{\d S_L}{\d \phi_{L}}.
\ee
On the other hand, for the energy-momentum tensor one may directly use the formulas \eqref{eq:euclT} and \eqref{eq:lorT}, where now $S_L$ and $S_E$ are the on-shell bulk action. It was shown in \cite{Skenderis:2008dg} that these expressions, with in particular the above choice of signs, lead to continuous holographic expressions for the one-point functions. For example, in the case of three-dimensional Einstein gravity one finds:
\be
\label{eq:einsteinvev}
\vev{T_{ij}} = \frac{1}{4 G_N} (g_{(2)ij} + \hf g_{(0)ij} R[g_{(0)}]),
\ee
independently of the metric signature. In this expression $g_{(0)ij}$ and $g_{(2)ij}$ the leading and subleading terms in the Fefferman-Graham expansion \eqref{eq:fgexpansion}. 
Similarly, for a scalar operator $\op$ dual to a bulk scalar field $\Phi$ one finds that:
\be
\label{eq:scalarvev}
\vev{\op} = -(2\Delta -d) \phi_{(2\Delta -d)}
\ee
with $\phi_{(2\Delta -d)}$ the coefficient of order $z^{\Delta}$ in the radial expansion \eqref{eq:expansionphi}. Again, with the above conventions the formula \eqref{eq:scalarvev} holds both in Lorentzian and in Euclidean signature \cite{Skenderis:2008dg}.

\section{Linearized equations of motion in global coordinates}
\label{sec:applinearizedeom}
In this appendix we will present the linearized equations in global coordinates. The usual metric
\be
ds^2 = -\cosh^2(r) dt^2 + \sinh^2(r) d\phi^2 + dr^2
\ee
can be put in the Fefferman-Graham form \eqref{eq:fgmetric} by defining
\be
\r = 4 e^{- 2 r},
\ee
after which we obtain:
\be
ds^2 = -\frac{1}{\r}\Big(1 + \hf \r + \frac{1}{16}\r^2\Big) dt^2 + \frac{1}{\r}\Big(1 - \hf \r + \frac{1}{16}\r^2\Big) d\phi^2 + \frac{d\r^2}{4 \r^2}.
\ee
These coordinates cover all of $AdS$ and are thus global coordinates.
Notice that $\del_k g_{ij} = 0$ and therefore $\G_{ij}^k[g] = 0$ (which of course does \emph{not} imply that $\delta \G_{ij}^k$ vanishes in the linearized equations). We also find that:
\be
(g' g^{-1}g')_{ij} = 2 g''_{ij}; \qquad g''_{ij} - \hf \tr(g^{-1}g') g'_{ij} = f(\r) g_{ij}; \qquad  \tr(g^{-1}g') = - 2 \r f(\r),
\ee
with
\be
f(\r) = \frac{2}{16 - \r^2},
\ee
which we use to simplify the formulas below. In the expressions below traces are implicitly taken with the aid of $g^{-1}$, that is we write $\tr(g')$ where before we wrote $\tr(g^{-1}g')$.

The linearized $(ij)$ equation of motion \eqref{eq:ijeom} becomes:
\begin{align}
& - h''_{ij} - \r f(\r) h'_{ij} + f(\r) h_{ij} +  g_{ij}\Big[\hf \tr(h'') - \hf \tr(h g^{-1}g'') + \r f(\r) (\tr(h') - \tr(g' g^{-1}h) ) \Big] \nn \\
&+ \hf g'_{ij} \Big[ \tr(h') - \tr(g'g^{-1}h) \Big] 
+ \frac{1}{\m}\e_i^{\phantom{i}k} \Big[ \qt \del_k \del^l h'_{lj} - \qt (g^{-1}g')^c_j
[\del_k \del^l h_{lc} - \hf \del_k \del_c \tr(h) ] + (j \leftrightarrow k) \Big] \nn \\
&+ \frac{1}{\m}\e_i^{\phantom{i}k} \Big[ \qt \del_k \del_j \tr(g'g^{-1}h) - \hf \del_k \del_j \tr(h') + 2 \r h'''_{jk}
+3 (1 + \r^2 f(\r)) [h''_{jk} + \r f(\r) h'_{jk} - f(\r)  h_{kj}] \Big] \nn \\
&+ \frac{1}{\m}\e_i^{\phantom{i}k} g'_{jk} \Big[
- \frac{3}{2}(1 + \r^2 f(\r))  [ \tr(h') - \tr(h g^{-1}g')]
- \frac{7}{2} \r  [\tr(h'') + \tr(hg^{-1}g'') - \tr(h' g^{-1}g') ]
\Big] \nn \\ &+ (i \leftrightarrow j) = 0,
\end{align}
The linearized version of the $(\r i)$ equation given in \eqref{eq:secondrijkl} becomes:
\begin{align}
& 2 \r\del^k h''_{ik} + (1 + 4 \r^2 f(\r)) \del^k h'_{ik} + \m \e^{jk}\del_k h'_{ij}
- \hf \m \e^{jk} (g^{-1}g')^l_j
 (\del_k h_{il} + \del_i h_{kl} - \del_l h_{ik})\nn\\
& - \del_i \Big[ \r \tr(h' g^{-1}g') + [1 + 4 \r^2 f(\r)] \tr(h') - [\hf + 2  \r^2 f(\r) ]\tr(g' g^{-1}h) - \r \tr(g'' g^{-1}h) \Big]\nn\\
&+ (g^{-1}g')^k_i \Big[\r \del^l h'_{kl} - 2\r \del_k \tr(h') - \frac{3}{2} \r \del_k \tr(h g^{-1}g') - [1 + 4 \r^2 f(\r)] [\del^l h_{kl} - \hf \del_k \tr(h) ]  \Big] \nn\\
& - 2 \r(g^{-1}g'')_i^k [2 \del^l h_{kl} - \del_k \tr(h) ] = 0
\end{align}
and the $(\r\r)$ equation results in:
\begin{align}
&- \tr(h'') + \tr(h' g^{-1}g') - \tr(h g^{-1}g'') + \frac{1}{2\m} \e^{ij} \Big[ \del_i \del^m h'_{mj} - (g^{-1}g')^c_j(\del_i \del^m h_{mc} - \hf \del_i \del_c \tr(h) )\nn\\
&\qquad \qquad + 2 \r (h'g^{-1}g'')_{ij}- 2 \r(g' g^{-1}h g^{-1}g'')_{ij} + 2 \r (g' g^{-1}h'')_{ij}
\Big] = 0.
\end{align}

\section{Some results from LCFT}
\label{sec:applcft}

A logarithmic conformal field theory (LCFT) is a conformal field theory in which logarithmic structure arises in
the operator product expansion. Such logarithmic structure arises when there are fields with degenerate
scaling dimensions having a Jordan block structure; in any logarithmic conformal field theory one of these degenerate
fields becomes a zero norm state coupled to a logarithmic partner. In what follows we will be interested in
the simplest situation, in which two operators become degenerate and form a logarithmic pair, denoted by $(C,D)$.
If the operator $C$ becomes a zero norm state, the two point functions for this logarithmic pair have the structure:
\begin{eqnarray}
\< C(z,\bar{z}) C(0) \> &=& 0; \qquad \< C(z,\bar{z}) D (0,0) \> = \frac{b_D}{2z^{2h_L} \bar{z}^{2 h_R}}; \label{eq:Jordan}\\
\< D(z, \bar{z}) D(0,0) \> &=& \frac{1}{z^{2h_L} \bar{z}^{2 h_R}}
\left [ - b_{D} \log m^2 |z|^2  + B_{D} \right ], \nn
\end{eqnarray}
where the conformal weights of both operators are $(h_L, h_R)$.
The constant $B_{D}$ may be removed by the redefinition $D \rightarrow D - B_{D} C/b_{D}$ but $b_{D}$ 
has an invariant meaning and is a characteristic of the LCFT.
One can easily generalize these formulas to the
case when there are $n$ degenerate fields and the Jordan cell is given by an $n \times n$ matrix, in which case the maximal
power of the logarithm will be $\log^n |z|$.

In the current context we are interested in the case where the
conformal field theory becomes logarithmic as $c_L \rightarrow 0$
and one of the logarithmic pair is the holomorphic
stress energy tensor.  
There are several distinct approaches to
taking such limits, see \cite{Flohr:2005dr} for a review, but the
limit relevant for us was discussed in Kogan and Nichols
\cite{Kogan:2002mg}.  The following is a slightly modified version of
the discussion in that paper, in which we take the limit $c_L
\rightarrow 0$ only in the holomorphic sector.

Consider a conformal field theory with central charges $(c_L, {c}_R)$ and holomorphic/anti-holomorphic
stress energy tensors $(T(z), \bar{T}( \bar{z}))$ respectively, such that
\be
\< T(z) T(0) \> = \frac{c_L}{2 z^4}; \qquad
\< \bar{T} (\bar{z}) \bar{T} (0) \> = \frac{c_R}{2 \bar{z}^4}.
\ee
Let $V(z, \bar{z})$ be a primary field of dimensions $(h_L, h_R)$, normalized as
\be
\< V(z, \bar{z}) V(0,0) \> = \frac{A}{z^{2h_L} \bar{z}^{2 h_R}}.
\ee
If $T$ is the only $h_L=2$ field present (and $\bar{T}$ is the only ${h}_R = 2$ field),
then the OPE for $V(z, \bar{z})$ is of the form
\be
 V(z, \bar{z}) V(0, 0) \sim \frac{A}{z^{2h_L} \bar{z}^{2 {h}_R}}
\left [1 + \frac{2h_L}{c_L} z^2 T(0) + \frac{2 h_R}{{c}_R} \bar{z}^2 \bar{T} (0) + \cdots \right ]
\ee
where the ellipses denote operators of higher dimension.

Consider now the limit $c_L \rightarrow 0$ with ${c}_R$ finite: if $A$ remains finite in this limit then
the OPE is not well-defined. Suppose that as $c_L$ approaches zero then there is another field $X$ with
dimension $(2 + \lambda, \lambda)$ which approaches $(2,0)$; suppose also that its normalization is such that this field contributes to the OPE as
\be \label{v-ope}
V(z, \bar{z}) V(0,0) \sim \frac{A}{z^{2h_L} \bar{z}^{2 {h}_R}}
\left [1 + \frac{2h_L}{c_L} z^2 T(0) + \frac{2h_R}{c_R} z^{2 + \lambda} \bar{z}^{\lambda} X(0,0) + \cdots \right ].
\ee
Let the two-point function of $X$ be given by:
\be
\< X(z, \bar{z}) X(0,0) \> = \frac{B(\lambda)}{z^{4 + 2 \lambda} \bar{z}^{2 \lambda}},
\ee
whilst $\< T(z_1) X(z_2, \bar{z}_2 ) \>$ vanishes as they have different dimensions.
Now let us define a new field $t(z, \bar{z})$ via
\be
t = - \frac{1}{\lambda} T - \frac{1}{\lambda} X.
\ee
In this way the OPE (\ref{v-ope}) is rendered well-defined as $c_L \rightarrow 0$:
\be
V(z, \bar{z}) V(0,0) \sim \frac{A}{z^{2h_L} \bar{z}^{2 {h}_R}}
\left [1 + \frac{2h_L}{b} z^2 \big[ t(0,0) + T(0) \log (m^2 |z|^2) \big] + \cdots \right ],
\ee
provided the parameter $b$, defined as
\be
\label{eq:b}
b \equiv - \lim_{c_L \rightarrow 0} \frac{c_L}{\lambda(c_L)} = - \frac{1}{\lambda'(0)},
\ee
is finite. As $c_L \rightarrow 0$ the two point functions of the pair $(T,t)$ become:
\bea
\< T(z) T(0) \> &=& 0; \qquad \< T(z) t (0,0) \> = \frac{b}{2z^4}; \label{eq:tTtcorr}\\
\< t(z, \bar{z}) t(0,0) \> &=& \frac{1}{z^4}
\lim_{c_L \to 0} \left [ - \frac{b}{2\lambda} + \frac{B}{\lambda^2} - 2 \lambda B \log (m^2 |z^2|) +
\cdots \right ]. \nn
\eea
For this to be well-defined as $c_L \rightarrow 0$,
\be
B(c_L) = \frac{b \lambda}{2} + B_m \lambda^2 + {\cal O}(\lambda^3),
\ee
and therefore
\be
\< t(z, \bar{z}) t(0,0) \> = \frac{B_m - b \log (m^2 |z|^2)}{z^4}. \label{tt}
\ee
The logarithmic pair $(T,t)$ thus indeed has the anticipated two-point function structure given in (\ref{eq:Jordan}).
We are interested in the case where $c_R \neq 0$, and thus there is no such degeneration
in the anti-holomorphic sector. Note that
\be
\< \bar{T} (\bar{z}) t(0,0) \> = 0.
\ee
Recall that the constant $B_m$ can be changed by a redefinition of $t$; choosing
$t \rightarrow t - B_m T/b$ removes the non-logarithmic term in the two point
function (\ref{tt}).

\section{Warped AdS}
\label{app:warped}
The metric of global AdS$_3$ can be written in `warped' form as:
\be
ds^2 =  - \cosh^2(\s) d\tau^2 + \frac{1}{4}d\s^2 + (du + \sinh (\s) d\tau)^2
\ee
We can define:
\be
z = 2 e^{-\s/2} \qquad \qquad \s = 2 \log(z/2)
\ee
after which the metric becomes:
\be
ds^2 =  \frac{dz^2}{z^2} - d\tau^2 + du^2 + (\frac{4}{z^2} - \frac{z^2}{4}) du d\tau.
\ee
In this coordinate system it is manifest that this metric is conformally compact. Namely, $z$ can be used as the defining function: in agreement with the discussion in section \ref{sec:aladsspacetimes}, $z$ has a single zero at $z=0$ and the metric:
\be
z^2 ds^2 = dz^2 + 4 du d\tau + \ldots
\ee
is a non-degenerate three-dimensional metric that extends smoothly to $z=0$.

On the other hand, the metric of spacelike warped AdS can be written as:
\be
\label{eq:warpedads}
ds^2 =  \Big( - \cosh^2(\s) (\n^2 + 3) + 4 \n^2 \sinh^2(\s) \Big) d\tau^2 + \frac{d\s^2}{\n^2 + 3} + 4 \n^2 du^2 + 8 \n^2 \sinh(\s) du d\tau,
\ee
with $\n = \m/3$. After the coordinate transformation:
\be
\s = - \sqrt{\n^2 + 3} \log(z)
\ee
it becomes asymptotically of the form:
\be
ds^2 = \frac{dz^2}{z^2} + 3 (\n^2 - 1) z^{- 2 \sqrt{\n^2 + 3}}d\tau^2  + 8 \n^2 z^{- \sqrt{\n^2 + 3}}  du d\tau +  \ldots
\ee
As $z\to 0$, we find that the terms have a different pole structure and therefore this metric cannot be made regular by multiplication with the usual defining function $z$, unless $\n^2 = 1$ (which is AdS). Furthermore, the leading term in the induced metric at slices of constant $z$ is proportional to $d\tau^2$ and so it is degenerate.
 Thus the spacetime with metric \eqref{eq:warpedads} is not conformally compact. Notice that the same conclusion holds for any spacetime whose metric asymptotes to \eqref{eq:warpedads}.

For timelike warped AdS the metric has the form:
\be
ds^2 =  \Big( \cosh^2(\s) (\n^2 + 3) - 4 \n^2 \sinh^2(\s) \Big) du^2 + \frac{d\s^2}{\n^2 + 3} - 4 \n^2 d\tau^2 - 8 \n^2 \sinh(\s) du d\tau.
\ee
This is just spacelike warped AdS with the replacement $\tau \to i u$ and $u \to i\tau$ and we can immediately draw the same conclusions as for spacelike warped AdS.

For null warped AdS the metric is given by:
\be
ds^2 = \frac{dz^2}{z^2} + \frac{d u dv}{z^2} \pm \frac{du^2}{z^4},
\ee
which is a solution of TMG with $\m = 3$ or $\n = 1$. We again find a different pole structure for the different terms, as well as a singular leading-order term in the induced metric on slices of constant $z$. Again, no good defining function exists that makes the three-dimensional metric regular on the slice $z=0$ and this manifold is not conformally compact.

\providecommand{\href}[2]{#2}\begingroup\raggedright\endgroup


\begin{thebibliography}{10}

\bibitem{Deser:1982vy}
S.~Deser, R.~Jackiw, and S.~Templeton, ``{Three-Dimensional Massive Gauge
  Theories},''
\href{http://dx.doi.org/10.1103/PhysRevLett.48.975}{{\em Phys. Rev. Lett.} {\bf
  48} (1982)  975--978}.

\bibitem{Deser:1981wh}
S.~Deser, R.~Jackiw, and S.~Templeton, ``{Topologically massive gauge
  theories},''
\href{http://dx.doi.org/10.1016/0003-4916(82)90164-6}{{\em Ann. Phys.} {\bf
  140} (1982)  372--411}.

\bibitem{Li:2008dq}
W.~Li, W.~Song, and A.~Strominger, ``{Chiral Gravity in Three Dimensions},''
  \href{http://dx.doi.org/10.1088/1126-6708/2008/04/082}{{\em JHEP} {\bf 04}
  (2008)  082},
\href{http://arxiv.org/abs/0801.4566}{{\tt arXiv:0801.4566 [hep-th]}}.

\bibitem{Carlip:2008jk}
S.~Carlip, S.~Deser, A.~Waldron, and D.~K. Wise, ``{Cosmological Topologically
  Massive Gravitons and Photons},''
  \href{http://dx.doi.org/10.1088/0264-9381/26/7/075008}{{\em Class. Quant.
  Grav.} {\bf 26} (2009)  075008},
\href{http://arxiv.org/abs/0803.3998}{{\tt arXiv:0803.3998 [hep-th]}}.

\bibitem{Grumiller:2008qz}
D.~Grumiller and N.~Johansson, ``{Instability in cosmological topologically
  massive gravity at the chiral point},''
  \href{http://dx.doi.org/10.1088/1126-6708/2008/07/134}{{\em JHEP} {\bf 07}
  (2008)  134},
\href{http://arxiv.org/abs/0805.2610}{{\tt arXiv:0805.2610 [hep-th]}}.

\bibitem{Park:2008yy}
M.-i. Park, ``{Constraint Dynamics and Gravitons in Three Dimensions},''
  \href{http://dx.doi.org/10.1088/1126-6708/2008/09/084}{{\em JHEP} {\bf 09}
  (2008)  084},
\href{http://arxiv.org/abs/0805.4328}{{\tt arXiv:0805.4328 [hep-th]}}.

\bibitem{Grumiller:2008pr}
D.~Grumiller, R.~Jackiw, and N.~Johansson, ``{Canonical analysis of
  cosmological topologically massive gravity at the chiral point},''
\href{http://arxiv.org/abs/0806.4185}{{\tt arXiv:0806.4185 [hep-th]}}.

\bibitem{Carlip:2008eq}
S.~Carlip, S.~Deser, A.~Waldron, and D.~K. Wise, ``{Topologically Massive AdS
  Gravity},'' \href{http://dx.doi.org/10.1016/j.physletb.2008.07.057}{{\em
  Phys. Lett.} {\bf B666} (2008)  272--276},
\href{http://arxiv.org/abs/0807.0486}{{\tt arXiv:0807.0486 [hep-th]}}.

\bibitem{Carlip:2008qh}
S.~Carlip, ``{The Constraint Algebra of Topologically Massive AdS Gravity},''
  \href{http://dx.doi.org/10.1088/1126-6708/2008/10/078}{{\em JHEP} {\bf 10}
  (2008)  078},
\href{http://arxiv.org/abs/0807.4152}{{\tt arXiv:0807.4152 [hep-th]}}.

\bibitem{Giribet:2008bw}
G.~Giribet, M.~Kleban, and M.~Porrati, ``{Topologically Massive Gravity at the
  Chiral Point is Not Chiral},''
  \href{http://dx.doi.org/10.1088/1126-6708/2008/10/045}{{\em JHEP} {\bf 10}
  (2008)  045},
\href{http://arxiv.org/abs/0807.4703}{{\tt arXiv:0807.4703 [hep-th]}}.

\bibitem{Blagojevic:2008bn}
M.~Blagojevic and B.~Cvetkovic, ``{Canonical structure of topologically massive
  gravity with a cosmological constant},''
\href{http://arxiv.org/abs/0812.4742}{{\tt arXiv:0812.4742 [gr-qc]}}.

\bibitem{Li:2008yz}
W.~Li, W.~Song, and A.~Strominger, ``{Comment on 'Cosmological Topological
  Massive Gravitons and Photons'},''
\href{http://arxiv.org/abs/0805.3101}{{\tt arXiv:0805.3101 [hep-th]}}.

\bibitem{AyonBeato:2004fq}
 E.~Ayon-Beato and M.~Hassaine,
 ``pp waves of conformal gravity with self-interacting source,''
 Annals Phys.\  {\bf 317}, 175 (2005)
 [arXiv:hep-th/0409150].

\bibitem{AyonBeato:2005qq}
 E.~Ayon-Beato and M.~Hassaine,
 ``Exploring AdS waves via nonminimal coupling,''
 Phys.\ Rev.\  D {\bf 73}, 104001 (2006)
 [arXiv:hep-th/0512074].

\bibitem{Brown:1986nw}
J.~D. Brown and M.~Henneaux, ``{Central Charges in the Canonical Realization of
  Asymptotic Symmetries: An Example from Three-Dimensional Gravity},''
\href{http://dx.doi.org/10.1007/BF01211590}{{\em Commun. Math. Phys.} {\bf 104}
  (1986)  207--226}.

\bibitem{Strominger:2008dp}
A.~Strominger, ``{A Simple Proof of the Chiral Gravity Conjecture},''
\href{http://arxiv.org/abs/0808.0506}{{\tt arXiv:0808.0506 [hep-th]}}.

\bibitem{Maloney:2009ck}
A.~Maloney, W.~Song, and A.~Strominger, ``{Chiral Gravity, Log Gravity and
  Extremal CFT},''
\href{http://arxiv.org/abs/0903.4573}{{\tt arXiv:0903.4573 [hep-th]}}.

\bibitem{Carlip:2009ey}
S.~Carlip, ``{Chiral Topologically Massive Gravity and Extremal B-F Scalars},''
\href{http://arxiv.org/abs/0906.2384}{{\tt arXiv:0906.2384 [hep-th]}}.

\bibitem{Grumiller:2008es}
D.~Grumiller and N.~Johansson, ``{Consistent boundary conditions for
  cosmological topologically massive gravity at the chiral point},''
  \href{http://dx.doi.org/10.1142/S0218271808014096}{{\em Int. J. Mod. Phys.}
  {\bf D17} (2009)  2367--2372},
\href{http://arxiv.org/abs/0808.2575}{{\tt arXiv:0808.2575 [hep-th]}}.

\bibitem{Henneaux:2009pw}
M.~Henneaux, C.~Martinez, and R.~Troncoso, ``{Asymptotically anti-de Sitter
  spacetimes in topologically massive gravity},''
\href{http://arxiv.org/abs/0901.2874}{{\tt arXiv:0901.2874 [hep-th]}}.

\bibitem{Skenderis:2002wp}
K.~Skenderis, ``Lecture notes on holographic renormalization,'' {\em Class.
  Quant. Grav.} {\bf 19} (2002)  5849--5876,
\href{http://arxiv.org/abs/hep-th/0209067}{{\tt hep-th/0209067}}.

\bibitem{Henningson:1998gx}
M.~Henningson and K.~Skenderis, ``{The holographic Weyl anomaly},'' {\em JHEP}
  {\bf 07} (1998)  023,
\href{http://arxiv.org/abs/hep-th/9806087}{{\tt arXiv:hep-th/9806087}}.

\bibitem{Henningson:1998ey}
M.~Henningson and K.~Skenderis, ``{Holography and the Weyl anomaly},'' {\em
  Fortsch. Phys.} {\bf 48} (2000)  125--128,
\href{http://arxiv.org/abs/hep-th/9812032}{{\tt arXiv:hep-th/9812032}}.

\bibitem{Balasubramanian:1999re}
V.~Balasubramanian and P.~Kraus, ``{A stress tensor for Anti-de Sitter
  gravity},'' \href{http://dx.doi.org/10.1007/s002200050764}{{\em Commun. Math.
  Phys.} {\bf 208} (1999)  413--428},
\href{http://arxiv.org/abs/hep-th/9902121}{{\tt arXiv:hep-th/9902121}}.

\bibitem{deHaro:2000xn}
S.~de~Haro, S.~N. Solodukhin, and K.~Skenderis, ``{Holographic reconstruction
  of spacetime and renormalization in the AdS/CFT correspondence},''
  \href{http://dx.doi.org/10.1007/s002200100381}{{\em Commun. Math. Phys.} {\bf
  217} (2001)  595--622},
\href{http://arxiv.org/abs/hep-th/0002230}{{\tt arXiv:hep-th/0002230}}.

\bibitem{Skenderis:2000in}
K.~Skenderis, ``{Asymptotically Anti-de Sitter spacetimes and their stress
  energy tensor},'' {\em Int. J. Mod. Phys.} {\bf A16} (2001)  740--749,
\href{http://arxiv.org/abs/hep-th/0010138}{{\tt hep-th/0010138}}.

\bibitem{Papadimitriou:2005ii}
I.~Papadimitriou and K.~Skenderis, ``{Thermodynamics of asymptotically locally
  AdS spacetimes},'' {\em JHEP} {\bf 08} (2005)  004,
\href{http://arxiv.org/abs/hep-th/0505190}{{\tt hep-th/0505190}}.

\bibitem{Graham:1999jg}
C.~R. Graham, ``{Volume and area renormalizations for conformally compact
  Einstein metrics},''
\href{http://arxiv.org/abs/math/9909042}{{\tt arXiv:math/9909042}}.

\bibitem{Anderson:2004yi}
M.~T. Anderson, ``Geometric aspects of the {AdS/CFT} correspondence,''
\href{http://arxiv.org/abs/hep-th/0403087}{{\tt hep-th/0403087}}.

\bibitem{us}
K.~Skenderis and B.~C. van Rees, ``{Real-time gauge/gravity duality},''
  \href{http://dx.doi.org/10.1103/PhysRevLett.101.081601}{{\em Phys. Rev.
  Lett.} {\bf 101} (2008)  081601},
\href{http://arxiv.org/abs/0805.0150}{{\tt arXiv:0805.0150 [hep-th]}}.

\bibitem{Skenderis:2008dg}
K.~Skenderis and B.~C. van Rees, ``{Real-time gauge/gravity duality:
  Prescription, Renormalization and Examples},''
\href{http://arxiv.org/abs/0812.2909}{{\tt arXiv:0812.2909 [hep-th]}}.

\bibitem{FeffermanGraham}
C.~Fefferman and C.~Graham, ``{Conformal Invariants},'' {\em Elie Cartan et les
  Math{\'e}matiques d'aujourd'hui (Asterisque 95)} (1985)  .

\bibitem{Berg:2001ty}
M.~Berg and H.~Samtleben, ``{An exact holographic RG flow between 2d conformal
  fixed points},'' {\em JHEP} {\bf 05} (2002)  006,
\href{http://arxiv.org/abs/hep-th/0112154}{{\tt arXiv:hep-th/0112154}}.

\bibitem{Bianchi:2001de}
M.~Bianchi, D.~Z. Freedman, and K.~Skenderis, ``{How to go with an RG flow},''
  {\em JHEP} {\bf 08} (2001)  041,
\href{http://arxiv.org/abs/hep-th/0105276}{{\tt arXiv:hep-th/0105276}}.

\bibitem{Bianchi:2001kw}
M.~Bianchi, D.~Z. Freedman, and K.~Skenderis, ``{Holographic
  renormalization},'' {\em Nucl. Phys.} {\bf B631} (2002)  159--194,
\href{http://arxiv.org/abs/hep-th/0112119}{{\tt arXiv:hep-th/0112119}}.

\bibitem{Kanitscheider:2006zf}
I.~Kanitscheider, K.~Skenderis, and M.~Taylor, ``{Holographic anatomy of
  fuzzballs},'' {\em JHEP} {\bf 04} (2007)  023,
\href{http://arxiv.org/abs/hep-th/0611171}{{\tt arXiv:hep-th/0611171}}.

\bibitem{Gubser:1998bc}
S.~S. Gubser, I.~R. Klebanov, and A.~M. Polyakov, ``{Gauge theory correlators
  from non-critical string theory},'' {\em Phys. Lett.} {\bf B428} (1998)
  105--114,
\href{http://arxiv.org/abs/hep-th/9802109}{{\tt hep-th/9802109}}.

\bibitem{Witten:1998qj}
E.~Witten, ``{Anti-de Sitter space and holography},'' {\em Adv. Theor. Math.
  Phys.} {\bf 2} (1998)  253--291,
\href{http://arxiv.org/abs/hep-th/9802150}{{\tt hep-th/9802150}}.

\bibitem{Anninos:2008fx}
D.~Anninos, W.~Li, M.~Padi, W.~Song, and A.~Strominger, ``{Warped AdS\_3 Black
  Holes},'' \href{http://dx.doi.org/10.1088/1126-6708/2009/03/130}{{\em JHEP}
  {\bf 03} (2009)  130},
\href{http://arxiv.org/abs/0807.3040}{{\tt arXiv:0807.3040 [hep-th]}}.

\bibitem{Solodukhin:2005ah}
S.~N. Solodukhin, ``{Holography with Gravitational Chern-Simons Term},''
  \href{http://dx.doi.org/10.1103/PhysRevD.74.024015}{{\em Phys. Rev.} {\bf
  D74} (2006)  024015},
\href{http://arxiv.org/abs/hep-th/0509148}{{\tt arXiv:hep-th/0509148}}.

\bibitem{Kraus:2005zm}
P.~Kraus and F.~Larsen, ``{Holographic gravitational anomalies},'' {\em JHEP}
  {\bf 01} (2006)  022,
\href{http://arxiv.org/abs/hep-th/0508218}{{\tt arXiv:hep-th/0508218}}.

\bibitem{Moussa:2003fc}
K.~A. Moussa, G.~Clement, and C.~Leygnac, ``{The black holes of topologically
  massive gravity},'' {\em Class. Quant. Grav.} {\bf 20} (2003)  L277--L283,
\href{http://arxiv.org/abs/gr-qc/0303042}{{\tt arXiv:gr-qc/0303042}}.

\bibitem{Bardeen:1984pm}
W.~A. Bardeen and B.~Zumino, ``{Consistent and Covariant Anomalies in Gauge and
  Gravitational Theories},''
\href{http://dx.doi.org/10.1016/0550-3213(84)90322-5}{{\em Nucl. Phys.} {\bf
  B244} (1984)  421}.

\bibitem{AlvarezGaume:1984dr}
L.~Alvarez-Gaume and P.~H. Ginsparg, ``{The Structure of Gauge and
  Gravitational Anomalies},''
\href{http://dx.doi.org/10.1016/0003-4916(85)90087-9}{{\em Ann. Phys.} {\bf
  161} (1985)  423}.

\bibitem{Hotta:2009zn}
K.~Hotta, Y.~Hyakutake, T.~Kubota, T.~Nishinaka, and H.~Tanida, ``{Left-Right
  Asymmetric Holographic RG Flow with Gravitational Chern-Simons Term},''
\href{http://arxiv.org/abs/0906.1255}{{\tt arXiv:0906.1255 [hep-th]}}.

\bibitem{Imbimbo:1999bj}
C.~Imbimbo, A.~Schwimmer, S.~Theisen, and S.~Yankielowicz, ``{Diffeomorphisms
  and holographic anomalies},''
  \href{http://dx.doi.org/10.1088/0264-9381/17/5/322}{{\em Class. Quant. Grav.}
  {\bf 17} (2000)  1129--1138},
\href{http://arxiv.org/abs/hep-th/9910267}{{\tt arXiv:hep-th/9910267}}.

\bibitem{Skenderis:1999nb}
K.~Skenderis and S.~N. Solodukhin, ``Quantum effective action from the
  {AdS/CFT} correspondence,'' {\em Phys. Lett.} {\bf B472} (2000)  316--322,
\href{http://arxiv.org/abs/hep-th/9910023}{{\tt hep-th/9910023}}.

\bibitem{Gurarie:2004ce}
V.~Gurarie and A.~W.~W. Ludwig, ``{Conformal field theory at central charge c =
  0 and two- dimensional critical systems with quenched disorder},''
\href{http://arxiv.org/abs/hep-th/0409105}{{\tt arXiv:hep-th/0409105}}.

\bibitem{Kogan:2002mg}
I.~I. Kogan and A.~Nichols, ``{Stress energy tensor in c = 0 logarithmic
  conformal field theory},''
\href{http://arxiv.org/abs/hep-th/0203207}{{\tt arXiv:hep-th/0203207}}.

\bibitem{Flohr:2005dr}
M.~Flohr and A.~Muller-Lohmann, ``{Notes on non-trivial and logarithmic CFTs
  with c = 0},'' {\em J. Stat. Mech.} {\bf 0604} (2006)  P002,
\href{http://arxiv.org/abs/hep-th/0510096}{{\tt arXiv:hep-th/0510096}}.

\bibitem{Bergshoeff:2009hq}
E.~A. Bergshoeff, O.~Hohm, and P.~K. Townsend, ``{Massive Gravity in Three
  Dimensions},'' {\em Phys. Rev. Lett.} {\bf 102} (2009)  201301,
\href{http://arxiv.org/abs/0901.1766}{{\tt arXiv:0901.1766 [hep-th]}}.

\end{thebibliography}
\end{document}